\documentstyle[12pt,epsfig,cite]{article}
\newlength{\dinwidth}
\newlength{\dinmargin}
\begin{document}
\newcommand{\be}{\begin{equation}}
\newcommand{\ee}{\end{equation}}
\newcommand{\ba}{\begin{eqnarray}}
\newcommand{\tdm}[1]{\mbox{\boldmath $#1$}}
\def\qqd{(q\bar q)_{\rm\small dipole}}

\titlepage
\begin{flushright}
{\sc TSL/ISV-2001-0253      \\
October 31, 2001}       \\
\end{flushright}
\begin{center}
\vspace*{2cm}
{\Large \bf Saturation model for two-photon interactions \\ at high energies}

\vspace*{1cm}
 N. T\^\i mneanu$^{a}$, J.\ Kwieci\'{n}ski$^{b}$,  L. Motyka$^{a,c}$
\vspace*{0.5cm}
\begin{center}
{$^{a}$ High Energy Physics, Uppsala University, Uppsala, Sweden}   \\
{$^{b}$ H.\ Niewodnicza\'{n}ski Institute of Nuclear Physics,
Krak\'{o}w, Poland} \\
{$^{c}$ Institute of Physics, Jagellonian University, Krak\'{o}w, Poland}\\

\end{center}
\end{center}
\vspace*{1.5cm}
\vspace*{2cm}
\begin{abstract}
We formulate and analyse a saturation model for the total $\gamma \gamma$
and $\gamma^{*} \gamma^{*}$ cross-sections and for the real photon structure
function $F_2^{\gamma}(x,Q^2)$. The model is based on a picture in which
the $\gamma^{*} \gamma^{*}$ total cross-section for arbitrary photon
virtualities is driven by the interaction of colour dipoles, into which the
virtual photons fluctuate. The cross-section describing this interaction is
assumed to satisfy the saturation property with the saturation radius taken
from the Golec-Biernat and W\"{u}sthoff analysis of the $\gamma^*p$
interaction at HERA. The model is supplemented by the QPM and non-pomeron
reggeon contributions. The model gives a very good description of the data on
the $\gamma \gamma$ total cross-section, on the photon structure function
$F_2^{\gamma}(x,Q^2)$ at low $x$ and on the $\gamma^* \gamma^*$
cross-section extracted from LEP double tagged events. Production of heavy
flavours in $\gamma \gamma$ collisions is also studied. Predictions of the
model for the very high energy range which will be probed at future linear
colliders are given.
\end{abstract}
\newpage

\section{Introduction}

The concept of parton density saturation in high energy scattering mediated
by strong interactions is certainly one of the most interesting recent
developments in QCD theory and phenomenology
\cite{GLR,MQ,NIKO,MUELLER,MUELLERS,COLLINS,BARTELS,VENUGOPALAN,SALAM,
LEVIN,BAL,WEIGERT,BRAUN,K,KSOL,KGBMW}.
It is well known, that the cross-sections for processes characterised by a
large scale $Q^2$ exhibit a steep power-like rise with the collision centre
of mass energy squared $W^2$, when $W^2\gg Q^2$.
Such a rise, attributed to the so called hard pomeron \cite{GLR,BFKL}, if
continued to arbitrarily large energies would eventually lead to break-down
of the $S$-matrix unitarity.
To avoid the apparent unitarity violation, effects related to
screening (shadowing) phenomena have to be considered.
They correspond to multiple bare pomeron exchanges and multi-pomeron
interactions, and tame the steep rise at high energies.
In the language of parton densities, the cross-section rises
as the number of partons grows in the target, due to gluon emissions
into the available rapidity interval, $Y \,\sim \, \ln (W^2/Q^2)$.
This growth may be described as evolution in rapidity,
with the evolution length~$Y$.
However, apart from creation of new partons a competing phenomenon
of gluon recombination occurs, which reduces the number of partons.
The recombination becomes increasingly important at high parton densities,
i.e.\ for a large evolution length $Y$.
This qualitative picture has a solid theoretical basis, well rooted in QCD
\cite{GLR,MQ,NIKO,MUELLER,MUELLERS,COLLINS,BARTELS,VENUGOPALAN,SALAM,
LEVIN,BAL,WEIGERT,BRAUN,K,KSOL}.\\

A very non-trivial feature which arises from those studies is that the
characteristic rapidity evolution length $Y$, for which the unitarity
corrections become important, depends on the hard scale $Q^2$. This
statement may be inverted, leading to a notion of the $Y$-dependent
saturation scale $Q_s(Y)$, the characteristic scale for the transition
between colour transparency and saturated cross-section regimes, at given
rapidity $Y$. This phenomenon has recently been thoroughly studied through
the measurement of $ep$ inelastic scattering at HERA within a broad range
of $Q^2$ varying from the DIS large $Q^2$ region down to the real
photoproduction limit $Q^2 \approx 0$ \cite{MANDY}. Here $Q^2 = -q^2$,
where $q$ is the four-momentum transfer in the process $e p \rightarrow
e^{\prime} X$. Since this process is largely controlled by one photon
exchange, $Q^2$ corresponds to the photon virtuality.
Measurement of $ep$ inelastic scattering permits determination of the
virtual photon -- proton total cross-section $\sigma_{\gamma^*p}(Q^2,W^2)$
for all virtual photon polarisation states, with $W$ being
the photon -- proton collision energy.\\

Golec-Biernat and W\"{u}sthoff (GBW) managed to fit these data in a model
incorporating the saturation property, with rapidity dependent saturation
scale \cite{KGBMW}. In this case the $\gamma^* p$ cross-section is
described in terms of the $q\bar q$ colour dipoles which the (virtual)
photon fluctuates to according to a known wave function.
The dipoles scatter off the proton with a cross-section
which exhibits the colour transparency and the saturation
property in the limit of a small dipole size, $r \ll 1/Q_s(Y)$,
and a large dipole size, $r \gg 1/Q_s(Y)$, respectively.
It is an encouraging result that the predictions of the simple model agree
well with all the large rapidity data ranging from photoproduction to large
$Q^2$. This model would therefore provide a description of the transition
between the soft and hard high energy scattering in QCD. The same model
explains also properties of the cross-section for hard diffraction at HERA,
in particular the ratio of the total to diffractive cross-section being
constant with energy \cite{KGBMWD}. \\

Thus, it is important to perform other tests of the GBW model which
probe its universality. Two virtual photon interactions at high energies
offer an ideal opportunity for such studies since virtualities of both
photons can vary, so that the properties of the model may be studied
more extensively. Beside that, in the GBW model, the photon wave
function is known, contrary to the proton wave function, so
the two-photon data may be used to constrain the dipole-dipole cross-section
itself.\\

There exist several models of two-photon interactions
which aim at describing the variation of the dynamics
depending on the photon virtualities
\cite{GLM,GALUGA,SSGG,DOSCH,DDR,KOLYA,GOTSMAN,BKS,BKKS}.
Most of those models combine the Vector Meson Dominance with the Parton
Model suitably extended to the region of low virtualities.
They do also usually rely on the Regge pole description of the high energy
behaviour of the total cross-sections. Some of the models describing the
total cross-sections of real photons explore the minijet production
mechanism \cite{FS,CGP}. The $\gamma^* \gamma^*$ interactions have also
been described within the dipole picture \cite{DOSCH,DDR,KOLYA}, but
possible saturation properties of the dipole-dipole cross-section have not
been studied so far. The saturation ansatz may be useful to better
understand the features of the available two-photon data and to formulate
some interesting predictions for the two-photon physics at future linear
colliders. \\

In this paper we construct a generalisation of the saturation model for the
two-photon case, compare its predictions with the experimental data and
discuss the implications. The content of our paper is the following. In the
next section we recall the GBW saturation model for $\gamma^*p$
scattering and present its formulation for
$\gamma^* \gamma^*$ high energy interactions.
We point out that by a suitable choice of the quark masses, the model can
be used to describe the total $\gamma \gamma$ cross-section for two real
photons, the $\gamma^* \gamma^*$ total cross-section for two virtual photons
measured in double tagged $e^+e^- \rightarrow e^+e^- + X$ events and the
photon structure function $F_2 ^{\gamma} (x,Q^2)$ of real (or virtual)
photon for low values of the Bjorken parameter $x$. We show that in the
region where the saturation effects become important the model gives
steeper dependence of the cross-section on the collision energy
than that obtained in the case of $\gamma^*p$ scattering.
In Section~3 we present comparison with the available experimental data on
$\sigma_{\gamma \gamma}$,
$\sigma_{\gamma^* \gamma^*}$, $F^{\gamma}_2(x,Q^2)$ at low $x$ and for the
cross-section describing heavy flavour production in $\gamma \gamma$ collisions.
Section 4 contains our predictions for the above quantities in the very high energy regime,
which can be available in future linear $ee$, $\gamma e$ or $\gamma \gamma$
colliders. Finally in Section~5 we present a summary of our results.

\section{The saturation model}
\subsection{The Golec-Biernat--W\"{u}sthoff model}

The study of the total virtual photon -- proton cross-section in the
high~$W$ limit and for $Q^2$ ranging from small to large values allows
probing the transition from large to short distance physics in high energy
scattering. Numerous analyses exist which study this transition \cite{MANDY,BBJKRMP}.
Among them, a very successful description is provided by
the saturation model developed by Golec-Biernat and W\"usthoff \cite{KGBMW},
in which the $\gamma^*p$ scattering is viewed upon as the scattering between
$\qqd$ and the proton.
The colour dipoles $\qqd$ represent virtual components of the photon in the
transverse plane (the plane transverse to the collision axis)
and their distribution in the photon can be obtained in
the perturbative framework.
%
The cross-section $\sigma_{i}^{\gamma^*p}(Q^2,W^2)$ 
for the transversely ($i=T$) and longitudinally ($i=L$) polarised
virtual photon is given by the following formula
\begin{equation}
\sigma_{i}^{\gamma^*p}(Q^2,W^2)\; = \;
\int_0^1 dz \int d^2{\tdm r}\;
|\Psi_{i}(z,{\tdm r})|^2 \, \hat \sigma(x,r^2),
\label{dipole1}
\end{equation}
where ${\tdm r}$ denotes the transverse separation between $q$ and $\bar q$
in the colour dipole, $z$ is the longitudinal momentum fraction of the
quark in the photon and $x$ is the Bjorken parameter, i.e. $x=Q^2/(2pq)$.
The cross-section $\hat \sigma(x,r^2)$ is the $\qqd$--proton total
cross-section, and $|\Psi_{i}(z,{\bf r})|^2$ denotes the photon wave
function squared and summed over the quark helicities, 
in the photon polarisation state indicated by~$i$.
The photon wave function is given by its quark flavour decomposition
\begin{equation}
 |\Psi_{i}(z,{\tdm r})|^2\; = \;\sum_f |\Psi_{i}^f(z,{\tdm r})|^2,
\label{psisum}
\end{equation}
and
\be
|\Psi_{i}^f(z,{\tdm r})|^2\; = \;
{6\alpha_{em}\over 4 \pi^2} e_f^2
\cases{
[z^2+(1-z)^2]\;\epsilon_f ^2 K_1^2 (\epsilon_{f}r)
+m_f^2\,K_0^2(\epsilon_{f}r), \qquad \mbox{for $i=T$},
\cr \cr
4 Q^2 z^2 (1-z)^2 \; K_0^2 (\epsilon_{f}r), \qquad\qquad \mbox{for $i=L$}, }
\label{psit}
\end{equation}
with
\begin{equation}
\epsilon_{f}^2 \; = \; z(1-z)Q^2+m_f^2,
\label{epsilon}
\end{equation}
where $e_f$ and $m_f$ denote the charge and mass of the quark of flavour $f$.
The functions $K_0$ and $K_1$ are the McDonald--Bessel functions.\\

Equation (\ref{dipole1}) is in fact equivalent to the $k_t$ (or high energy) factorisation,
which is the basic tool for calculating the observable quantities at low $x$ \cite{KTFAC}, and
the dipole-proton cross-section  $\hat \sigma(x,r^2)$ is related to the
unintegrated gluon distribution in the proton $f(x,k^2)$ \cite{BIALAS}
\be
\hat \sigma(x,r^2)\; = \; {4 \pi \alpha_s\over N_c} \int {d^2{\tdm k}\over k^4}[1-J_0(kr)]f(x,k^2).
\label{unglu}
\ee
In equation (\ref{unglu}), ${\tdm k}$ is the transverse momentum of the
gluon and $J_0(z)$ is the Bessel function. In the leading $\ln (1/x)$
approximation, the unintegrated gluon distribution $f(x,k^2)$ is given by
the solution of the BFKL equation which determines the `hard' pomeron in
the perturbative QCD \cite{BFKL}.
The exchange of the perturbative QCD pomeron  however violates unitarity at very large
energies. The novel feature of the saturation model is the incorporation of the unitarity
constraint on the level of the dipole-proton cross-section $\hat \sigma(x,r^2)$.
This is achieved by imposing the saturation property, i.e.
$\hat \sigma(x,r^2) \to \sigma_0$ for $r \gg R_0(x)$, where the saturation radius
$R_0(x)$ is a decreasing function with decreasing $x$
\be
R_0^2(x) \;\sim\; x^{\lambda},
\ee
and the cross-section $\sigma_0$ is independent of $x$.
In the limit $r \rightarrow 0$, it follows from the perturbative QCD calculations,
that the dipole cross-section exhibits the colour transparency property 
behaving as $\hat \sigma(x,r^2) \;\sim\; r^2/R_0^2(x)$
(modulo a logarithmic correction which can modify the $r$ dependence).
Those two properties are economically summarised
by the following simple parametrisation
\be
\hat \sigma(x,r^2)\; = \;\sigma_0 [1-\exp(-r^2/(4R_0^2(x))].
\label{sigsat}
\ee

\subsection{Generalisation of the GBW model for the two-photon cross-section}

The description of the $\gamma^* \gamma^*$ total cross-sections
within the formalism utilising the interaction of two colour dipoles, which
the virtual photons fluctuate into, has been discussed in detail in
\cite{DOSCH,DDR,KOLYA}. The dipole-dipole cross-sections were assumed to be
given by the Stochastic Vacuum Model \cite{DOSCH,DDR} or to follow from the
BFKL formalism \cite{KOLYA}. The novel feature of our approach is
incorporation of the saturation property of the dipole-dipole
cross-section. This makes it possible, in particular, to describe in a
unified way the variation of the energy dependence of the cross-sections
with the change of the virtualities of the photons.
In terms of the virtual photon four-momenta $q_1$ and $q_2$ we have
$Q^2 _{1,2}=-q_{1,2}^2$ and $W^2 = (q_1+q_2)^2$, see Fig.~\ref{diagram}.
The extension the saturation model to the case of
$\gamma^{*}(Q_1^2) \gamma^{*}(Q_2 ^2)$ cross-sections
for arbitrary virtualities $Q_{1,2}^2$ is given below.

\begin{figure}[t]
\begin{center}
\epsfig{width= 0.6\columnwidth,file=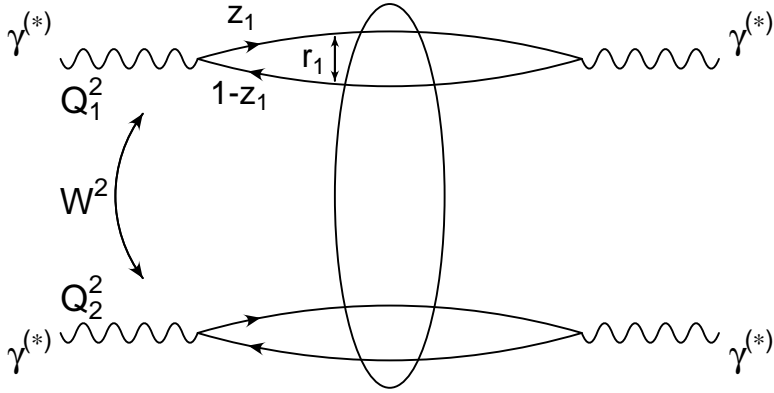}
\caption{\small\it The diagram illustrating the $\gamma^* \gamma^*$ interaction in the
dipole representation, see formula (\ref{master}).}
\label{diagram}
\end{center}
\end{figure}

A formula for the two-photon cross-section part coming from the exchange of
{\em gluonic} degrees of freedom reads \cite{DOSCH,DDR}
\be
\sigma^G_{ij}(W^2,Q_1^2,Q_2^2)\; = \;
\sum_{a,b=1}^{N_f} \int_0^1dz_1\int d^2 {\tdm r_1}|\Psi_i^a(z_1,{\tdm r_1})|^2
\int_0^1 dz_2\int d^2 {\tdm r_2}|\Psi_j^b(z_2,{\tdm r_2})|^2
\; \sigma^{dd}_{a,b}(\bar x_{ab},r_1,r_2) ,
\label{master}
\end{equation}
where
the indices $i,j$ label the polarisation states of the virtual photons,
i.e. $T$ or $L$. The wave functions $\Psi_i^a(z_k,{\bf r})$ are given by
equations (\ref{psit}), with $\epsilon_f^2$ defined by equation
(\ref{epsilon}) being replaced by $(\epsilon_f^k)^2=z_k(1-z_k) Q_k^2 +
m_f^2,\quad k=1,2\;$ and $\sigma^{dd}_{a,b}(\bar x_{ab},r_1,r_2)$ are the
dipole-dipole total cross-sections corresponding to their
different flavour content specified by the $a$ and $b$.

Inspired by the GBW simple choice for the dipole-proton cross-section,
we use the following parametrisation of the dipole-dipole cross-section
$\sigma_{a,b}(\bar x_{ab},r_1,r_2)$
\begin{equation}
\sigma^{dd}_{a,b}(\bar x_{ab},r_1,r_2)\; = \;\sigma_0^{a,b}\left[
1- \exp\left(-{r_{\rm\small eff}^2\over  4R_0^2(\bar x_{ab})}\right)
\right],
\label{sigmadd}
\end{equation}
where for $\bar x_{ab}$ we take the following expression symmetric in~$(1,2)$
\begin{equation}
\bar x_{ab} \; = \;{Q_1^2 + Q_2^2 +4m_a^2+4m_b^2\over W^2+Q_1^2+Q_2^2},
\label{barx}
\end{equation}
which allows an extension of the model down to the limit $Q_{1,2}^2=0$.
Note, that $\bar x_{ab}$  depends on the flavour of scattering quarks.
We use the same parametrisation of the saturation radius $R_0(\bar x)$
as that in equation (7) in \cite{KGBMW}, i.e.
\begin{equation}
R_0(\bar x)\; = \;{1\over Q_0} \left({\bar x\over x_0}\right)^{\lambda/2},
\label{r0}
\end{equation}
and adopt the same set of parameters defining this quantity as those in
\cite{KGBMW}. For the saturation value $\sigma_0^{a,b}$
of the dipole-dipole cross-section (cf.\ equation (\ref{sigmadd})) we set
\begin{equation}
\sigma_0^{a,b}\; = \;{2\over 3}\sigma_0,
\label{sigma0}
\end{equation}
where $\sigma_0$ is the same as that in \cite{KGBMW}.
For light flavours, equation (\ref{sigma0}) can be justified by the quark counting rule,
as the ratio between  the number of constituent quarks in a photon  and the
corresponding number of constituent quarks in the proton.
Following \cite{KGBMW},
we also use the same value of $\sigma_0^{a,b}$ for all flavours.\\

Three scenarios for $r_{\rm\small eff}(r_1,r_2)$ are considered:
\begin{enumerate}
\item $\displaystyle r^2_{\rm\small  eff}\; = \;{r_1^2r_2^2\over r_1^2+r_2^2}$,
\item $r^2_{\rm\small eff}\; = \;\min(r_1^2,r_2^2)$,
\item $r^2_{\rm\small  eff}\; = \;\min(r_1^2,r_2^2)[1+\ln(\max(r_1,r_2)/\min(r_1,r_2))]$.
\end{enumerate}
The first two cases are simple generalisations of the parametrisation
adopted in the case of the dipole-hadron scattering, i.e. we get
$r_{\rm\small eff}^2 \sim r_1^2$ ($r_{\rm\small eff}^2 \sim r_2^2$) in the
configurations $r_{2}^2 \gg r_1^2$ ($r_{1}^2 \gg r_2^2$). Case (3) is
motivated by the two-gluon exchange between the colour dipoles, giving
the following cross-section
\begin{equation}
\sigma_{dd}^{2g} \;\sim\; 2\int {dk^2\over k^4}[1-J_0(kr_1)][1-J_0(kr_2)]\; = \;
\min(r_1^2,r_2^2) [1+\ln(\max(r_1,r_2)/\min(r_1,r_2))].
\label{ggdd}
\end{equation}
In all three cases the dipole-dipole cross-section exhibits colour transparency,
i.e. $\sigma^{dd}_{a,b}(\bar x,r_1,r_2) \rightarrow~0$ for 
$r_{1} \rightarrow 0$ or $r_{2} \rightarrow 0$.\\

The formulae given above correspond to the `pomeron' contribution to the
$\gamma^* \gamma^*$ total cross-sections. This means that they represent
the exchange of gluonic degrees of freedom giving rise to the component
of cross-sections which, at high energies, does not decrease with
increasing energy. Equation (\ref{sigmadd}), defining the dipole-dipole
cross-section, similarly to equation~(\ref{sigsat}) for the dipole-proton
cross-section, interpolates between the `hard pomeron' at small transverse
separations $r_{1,2}$ and the `soft pomeron' at large transverse separations.
In more detail, for small values of $r_i$
(i.e.\ for $r_{\rm\small eff}^2 \ll 4 R_0^2(\bar x)$) one gets
\be
\sigma^{dd}_{a,b}(\bar x_{ab},r_1,r_2)\;\simeq\; \sigma_0^{a,b}
{ r_{\rm\small eff}^2 \over 4R_0^2(\bar x_{ab})} \;\sim\; \bar x_{ab}^{-\lambda},
\label{smallr}
\ee
which can be interpreted as the `hard pomeron' contribution if the parameter
$\lambda$ is identified with the `hard pomeron' intercept.
On the other hand for large dipole sizes we have
\be
\sigma^{dd}_{a,b}(\bar x_{ab},r_1,r_2)\;\sim\; \sigma_0^{a,b},
\ee
i.e.\ the cross-section is only slowly varying with the energy
in accordance with what is observed for the `soft pomeron'.
It should be emphasised that the structure of the saturation model
is different from that corresponding to two separate (i.e. `hard' and `soft')
pomeron contributions \cite{TWOP}. The soft pomeron in the saturation
model appears rather as an effect of unitarisation of the
exchange amplitude of the `hard' contribution, provided by multiple exchanges
and self-interaction of the `hard' pomerons.

\subsection{Scaling of the cross-sections}

The virtualities of the two photons can be arbitrary, thus the model
can describe the following three cases of physical and phenomenological
interest:
\begin{enumerate}
\item The case $Q_1^2 = Q_2^2 =0$ corresponding to the interaction of two real photons.
\item The case $Q_1^2 \sim Q_2^2$ (with large $Q_{1,2}^2$ ) corresponding to the interaction
  of two (highly) virtual photons.  The relevant cross-section can be extracted from
  the measurement of double tagged events
  $e^+e^- \rightarrow e^+e^- + \; {\rm\it hadrons}$.
\item The case $Q_1^2 \gg Q_2^2$  corresponding to probing the structure of virtual
  ($Q_2^2 > 0$)  or real ($Q_2^2 =0$) photon at small values of the Bjorken
  parameter $x=Q_1^2/(2q_1q_2)$.  For instance, the structure function
  $F_2^{\gamma}(x,Q^2)$ of the real photon ($Q_2^2=0, Q_1^2=Q^2$)
  is  related in the following way to the $\gamma^* \gamma$
  total cross-sections
\be
F_2 ^{\gamma}(x,Q^2)\; = \;
{Q^2\over 4 \pi^2 \alpha_{em}}
[\sigma_{T,T}(W^2,Q^2,Q_2^2=0) + \sigma_{L,T}(W^2,Q^2,Q_2^2=0)].
\label{fgg}
\ee
\end{enumerate}

Let us now examine the high energy limit in all three cases and compare
them with the $\gamma^{*}p$ case. We shall show that in the kinematical
configurations in which the saturation effects are important, the
two-photon cross-sections are more singular in the high energy limit than
the $\gamma^{*}p$ cross-section. To be precise, the high energy behaviour
of the total $\gamma^* \gamma^*$ cross-sections will be enhanced by
additional factors of the large logarithm $\ln[Q^2R^2_0(\bar x)]$. Also the
behaviour of the $\gamma \gamma$ total cross-section will be shown to be
enhanced by an additional power of $\ln(W^2/m_q ^2)$, i.e. the $\gamma \gamma$
total cross-section becomes steeper function of $W$ than the $\gamma p$
total cross-section, which seems to be confirmed experimentally. This
enhancement will be a direct consequence of the singular behaviour of the
photon wave function for small dipole sizes related to the point-like
component of the photon. 
For simplicity, in the analysis given below, we focus on
the contributions to the two-photon cross-sections of the light quarks 
$u,d$ and $s$  with the same mass $m_u = m_d = m_s  = m_q$.
The heavy flavour components exhibit the same general properties as the 
light flavour ones, but in the presently available kinematic
range, the transition to saturation regime may not be observed.

Let us first examine case (1) for the $\gamma \gamma$ total cross-section.
The dominant contribution to the integrals in equation (\ref{master}) comes
from the region
$4R_0^2(\bar x) \,\ll\, r_{1}^2 \,\ll\, r_2^2 \,\ll\, 1/m_q^2$,
$\, 4R_0^2(\bar x) \,\ll\, r_{2}^2 \,\ll\, r_1^2 \,\ll\, 1/m_q^2$,
where $\bar x = 8 m_q^2/W^2$ and $R_0(\bar x)$ is given by equation (\ref{r0}).
In this region the short distance approximation of the (transverse)
photon wave-function may be used
\be
|\Psi_T(z,{\bf r})|^2 \;\sim\; {1\over r^2},
\label{shdist}
\end{equation}
and the corresponding contribution to the total $\gamma \gamma$ cross-section is
\be
\sigma_{\gamma \gamma}(W^2) \;\sim\; \int_{4R_0^2(\bar x)}^{1/m_q^2}
{dr_2^2\over r_2^2}\int_{4R_0^2(\bar x)}^{r_2^2}
{dr_1^2\over r_1^2} \;\sim\; \ln^2 [4R_0^2(\bar x)m_q^2]\;\sim\; \ln^2(W^2/W_0^2).
\label{sggl2}
\ee
This should be compared with the $\gamma p$ total cross-section, where the
saturation model extended down to the photoproduction limit gives
$\sigma_{\gamma p}(W^2) \sim \ln(W^2/W_0^2)$. \\

Case (2) of the $\gamma^* \gamma^*$ cross-section in the configuration
$Q_1^2 \sim Q_2^2 \sim Q^2$, with $Q^2$ being large,  is
regarded as a very useful tool for probing the bare `hard' pomeron exchange amplitude
\cite{GSGSBFKL,JKLM}.
The short-distance approximation
(\ref{shdist}) of the photon wave function is now valid in the region
\be
r_{k}^2 \; z_{k}(1-z_{k}) \; Q_{k}^2 \;\ll\; 1, \qquad k=1,2.
\label{ranges}
\ee
The saturation model predicts different high energy behaviour of the
$\gamma^* \gamma^*$ cross-section  depending on whether
$Q^2\,>\, Q_s^2(\bar x)$ or $Q^2\,<\, Q_s^2(\bar x)$, 
with the saturation scale $Q_s^2(\bar x) \,\sim\, 1/R_0^2(\bar x)$.
Thus in the region $Q^2\,>\,Q_s^2(\bar x)$ we get 
(modulo logarithmic corrections)
\begin{equation}
\sigma_{\gamma^*\gamma^*}(W^2,Q_1^2\sim Q^2,Q_2^2\sim Q^2)
\;\sim\; {1\over Q^2 R_0^2(\bar x)},
\label{siglq2}
\ee
which exhibits the rise $W^{2\lambda}$ and $1/Q^2$ dependence,
characteristic for the hard pomeron exchange.
In the region $Q^2\,<\,Q_s^2(\bar x)$ the  
$\gamma^* \gamma^*$ cross-section has the saturation property, i.e.
\begin{enumerate}
\item[a)] the $1/Q^2$ behaviour is changed into a weakly varying function of $Q^2$;
\item[b)] the power-like $W^{2\lambda}$ behaviour is replaced by a moderately
          increasing function of $W^2$.
\end{enumerate}
The leading behaviour of $\sigma_{\gamma^*\gamma^*}$
for $Q^2R_0^2(\bar x) \,\ll\, 1$ comes from the `strongly ordered'
configurations in  integrals (\ref{master}) defining the
$\gamma^* \gamma^*$ total cross-sections
\be
{1\over z_k(1-z_k)Q^2} \;\gg\; r_2^2 \;\gg\; r_1^2 \;\gg\; 4R_0^2(\bar x),
\label{sohq}
\ee
with $k=1$ and 2.
From equations (\ref{master},\ref{shdist},\ref{sohq}) one obtains
\be
\sigma_{\gamma^*\gamma^*}(W^2,Q_1^2\sim Q^2,Q_2^2\sim Q^2)\;\sim\; \ln^2(Q^2 R_0^2(\bar x))\,
[1+{\cal O}(1/\ln(Q^2R_0^2(\bar x))].
\label{satgs}
\ee

Note that in the saturation model at high energies, the `pomeron' component of
the $\gamma^* \gamma^*$ total cross-section is, to a good approximation,
a function of only two variables $\tau_1=Q_1^2R_0^2(\bar x)$ and
$\tau_2=Q_2^2 R_0^2(\bar x)$. An analogous `geometric scaling' was found in the
DIS data \cite{GEOMETRIC}. A weak breaking of the scaling property in both cases
occurs due to the presence of quark masses.
For $Q_1^2 \sim Q_2^2 \sim Q^2$  (i.e. $\tau_1 \sim \tau_2 \sim \tau$),
the $\gamma^*\gamma^*$ cross-section exhibits the $1/\tau$ behaviour at
large $\tau$ (see eq.\ (\ref{siglq2})) and reaches the saturation limit
corresponding to a slowly varying function of $\tau$ for small values of $\tau$.
It should  be observed that the leading behaviour at small $\tau$ for
the $\gamma^* \gamma^*$ total cross-section (see equation (\ref{satgs}))
is more singular  than for the $\gamma^*p$ case, where $\sigma_{\gamma^*p}
\sim \ln(1/\tau)$.\\

Finally, in case (3) corresponding to probing the structure of the
real (or quasi-real) photons at low~$x$ and large~$Q^2$ we find
\begin{equation}
F^{\gamma}_2(x,Q^2) \; \sim \; x^{-\lambda},
\label{ff2}
\end{equation}
for $Q^2 \,>\, Q_s^2(x)$ and
 \begin{equation}
F^{\gamma} _2 (x,Q^2) \; \sim \; Q^2\, \ln^2 [Q^2 R_0^2(x)],
\label{ffsat2}
\end{equation}
for $Q_0^2 \,<\, Q^2 \,\ll\, Q_s^2(x)$.

\subsection{Non-pomeron contributions}

In order to get a complete description of $\gamma^* \gamma^*$ interactions, which
could be extended down to values of $W\sim 10\;{\rm GeV}$,  we should add to the
`pomeron' contribution defined by equation (\ref{master}) the non-pomeron reggeon and
QPM terms \cite{JKLM}.
The additional contributions are characterised by a decreasing
energy dependence, i.e.
$\sim 1/W^{2\eta}$ for the reggeon and $\sim 1/W^2$ (with $\ln W$ corrections)
for QPM.  The QPM contribution, represented by the quark box
diagrams, is well known and the cross-sections are given, for instance,
in \cite{BUD}.
The reggeon contribution represents a non-perturbative phenomenon
related to Regge trajectories of light mesons.
It is known mainly from fits to total hadronic cross-sections and to the proton
structure function $F_2$.
A state-of-art parametrisation of the reggeon exchange cross-section
in two-photon interactions is given by the following expression \cite{DDR}
\be
\sigma^R(W^2,Q_1^2,Q_2^2)\; = \; 4\pi^2\alpha_{em}^2\frac{A_2}{a_2}\left[{a_2^2\over
(a_2+Q_1^2)(a_2+Q_2^2)}\right]^{1-\eta}\left({W^2\over a_2}\right)^{-\eta}.
\label{reggeon}
\ee
Originally, it was set \cite{DDR}: $A_2 = 0.38$, $a_2=0.3\;{\rm GeV}^2$ and
$\eta=0.45$. However, those parameters were obtained with certain
assumptions concerning the pomeron exchange which are different from ours.
Therefore, it is legitimate to modify the parameters of~\cite{DDR}
while retaining its functional form. We have in particular found that in
order to get a good description of the data on the $\gamma \gamma$ total
cross-section in the `low' energy region, $W<10~{\rm GeV}$, one has to set
$\eta = 0.3$. This happens to be consistent with the recent observation
that the intercept of the $f_2$ trajectory, which contributes to the
two-photon cross-section, can be expected to be equal to 0.7 \cite{PVLF0}.
We fitted the other parameters, $A_2$ and $a_2$, to the data on two-photon
collisions. 

Finally, note, that the decomposition of the reggeon term into different 
photon polarisation states has not been specified. In our analysis we assume, 
that the reggeon couples only to transverse photons. This arbitrary assumption,
does not influence significantly our results for the studied observables. 

\subsection{Threshold corrections}

Strictly speaking, both the dipole model (accounting for the `pomerons')
and the Regge model of the total cross-section are formulated
in the high energy limit $x~\simeq~Q^2/W^2~\ll~1$.
When extending applicability of these models up to larger values of $x$, 
for instance $x \simeq 0.1$, threshold correction factors should be taken into account. 
Namely, the cross-section should vanish when $x \to 1$ as a power of
$1-x$. In the case of $\gamma^* p$ scattering, the form 
of the cross-section at $x \simeq 1$ is governed by the number of spectator quarks
$n_{\rm\small spect}$ in the proton which do not interact directly with the photon. 
To be precise, it follows from the dimensional-counting rules that 
for a subprocess with a given number of spectators, at $x \simeq 1$, 
the cross-section takes the form 
$\sigma_{\gamma^* p} (x,Q^2) \, \sim \, (1-x)^{2n_{\rm\tiny spect}-1}$
(where the $Q^2$ dependence is suppressed). 
A possible way to combine the small~$x$ dependence of the cross-section in the 
Regge model with the latter result is to include $(1-x)^{2n_{\rm\tiny spect}-1}$ 
as a multiplicative correction factor to the asymptotic cross-section from
the pomeron or a subleading reggeon  exchange \cite{DLspec}.
For the pomeron exchange in $\gamma^* p$ scattering one has $n_{\rm\small spect}=4$
and for the other reggeons ($f_2$ and $a_2$),  $n_{\rm\small spect}=2$.
It is clear, that a similar procedure may be applied for the saturation model 
as well.

In the case of two-photon collisions, one of the photons plays the role of 
the target, probed by the other photon. In the dipole representation, 
the number of valence quarks in the target photon equals two, to be compared with
three valence quarks in the proton. 
Thus, for the non-pomeron reggeons one has 
$n_{\rm\small spect}=1$ and for the dipole-dipole scattering component, which 
represents  the `pomeron' exchange, one obtains $n_{\rm\small spect}=3$.
Recall, that in order to extend the saturation model to describe real photons, 
we use the variable $\bar x_{ab}$ (see eq.~(\ref{barx})) instead of $x$. 
Also here, we represent the threshold correction factors using $\bar x_{ab}$.
Thus we multiply the reggeon term (\ref{reggeon}) by $(1-\bar x_{qq})$ with
$\bar x_{qq}$ obtained from eq.\ (\ref{barx}) with $m_a^2=m_b^2=m_q^2$,
that is defined by the light quark mass $m_q$.
For the dipole-dipole scattering cross-section, $\bar x_{ab}$ depends upon  
the flavour of quarks which span the dipoles. 
Hence, in our final formulae we multiply the dipole-dipole cross-section 
$\sigma^{dd}_{a,b}(\bar x_{ab},r_1,r_2)$, (see (\ref{sigmadd})) by the 
factor $(1-\bar x_{ab})^5$. 

\subsection{Final formulae}

For clarity, we collect the components of the two-photon cross-section 
presented above. The total $\gamma^*(Q_1^2)\gamma^*(Q_2^2)$ cross-section 
reads
\be
\sigma_{ij} ^{\rm\small tot} (W^2,Q_1^2,Q_2^2) \;=\;
\tilde\sigma_{ij} ^G (W^2,Q_1^2,Q_2^2) + 
\tilde\sigma ^R (W^2,Q_1^2,Q_2^2) \delta_{iT} \delta_{jT}+
\sigma_{ij} ^{\rm\tiny QPM} (W^2,Q_1^2,Q_2^2),
\label{total}
\ee
where
$
\tilde\sigma_{ij} ^G (W^2,Q_1^2,Q_2^2) 
$
is the gluonic component, corresponding to dipole-dipole scattering,
as in eq.\ (\ref{master}), but with the dipole-dipole cross-section
including the threshold correction factor 
\be
\tilde\sigma^{dd}_{a,b}(\bar x_{ab},r_1,r_2) \; = \;
     (1-\bar x_{ab})^5\, \sigma^{dd}_{a,b}(\bar x_{ab},r_1,r_2), 
\label{sigmaddth}
\ee
c.f.\ eq.\ (\ref{sigmadd}), and $\bar x_{ab}$ is given
by eq.\ (\ref{barx}).
The sub-leading reggeon contributes only to scattering of two 
transversely polarised photons and also contains a threshold
correction
\be
\tilde\sigma ^R (W^2,Q_1^2,Q_2^2) = (1-\bar{x})\, \sigma ^R (W^2,Q_1^2,Q_2^2),
\ee  
with 
\be
\bar x = {Q_1^2 + Q_2 ^2 + 8 m_q ^2 \over W^2 + Q_1^2 + Q_2 ^2}.
\ee
The third term $\sigma_{i,j} ^{\rm\tiny QPM} (W^2,Q_1^2,Q_2^2)$
is the standard QPM contribution and is taken from \cite{BUD}.

\section{Comparison to experimental data}

\subsection{Parameters of models}

In the comparison to the data we study three models, based on all cases
for the effective radius, as described in Section~2.2. We will refer to the
these models as Model 1, 2 and 3, corresponding to the choice of the
dipole-dipole cross-section.
Let us recall that we take without any modification the parameters
of the GBW model: $\sigma_0 = 29.13 $~mb, $x_0= 0.41 \cdot 10^{-4}$ and  $\lambda= 0.277$.
However, we fit the light quark mass to the two-photon data, since it is not
very well constrained by the GBW fit, as we explicitly verified.
On the other hand, the sensitivity of the choice
of the mass appears to be large for the two-photon total cross-section.
We find that the optimal values of the light quark ($u$, $d$ and $s$) masses
$m_q$ are 0.21, 0.23 and 0.30~GeV in Model 1,~2 and~3 correspondingly.
Also, the masses of the charm and bottom quark are tuned within the range
allowed by current measurements, to get the  optimal global description
in Model~1, $\displaystyle r^2_{\rm\small  eff}\; = {r_1^2r_2^2 / (r_1^2+r_2^2)}$,
which agrees best with data. For the charm quark we use $m_c = 1.3$~GeV
and for bottom $m_b = 4.5$~GeV.
Moreover, we re-fit $\eta$, $A_2$ and $a_2$ parameters in the 
reggeon term (\ref{reggeon}), which is legitimate because the `pomeron' term which we use is
different from this following from the model of two pomerons used in \cite{DDR}.
We find that the values $\eta = 0.3$, $A_2 = 0.26$ and $a_2 = 0.2\;{\rm GeV}^2$
give the  best description of data, when combined with the saturation model.
The values of masses listed above are consistently used also in the quark
box contribution (QPM). The Models, which we shall mention from now on, 
contain the saturation models described in Section 2, combined with
the reggeon and QPM contribution.

It should be stressed, that most of the data relevant for our
study were collected with the help of the comprehensive review~\cite{KZS}.

\subsection{The test case: the $\gamma p$ total cross-section}

\begin{figure}[t]
\begin{center}
\epsfig{width= 0.8\columnwidth,file=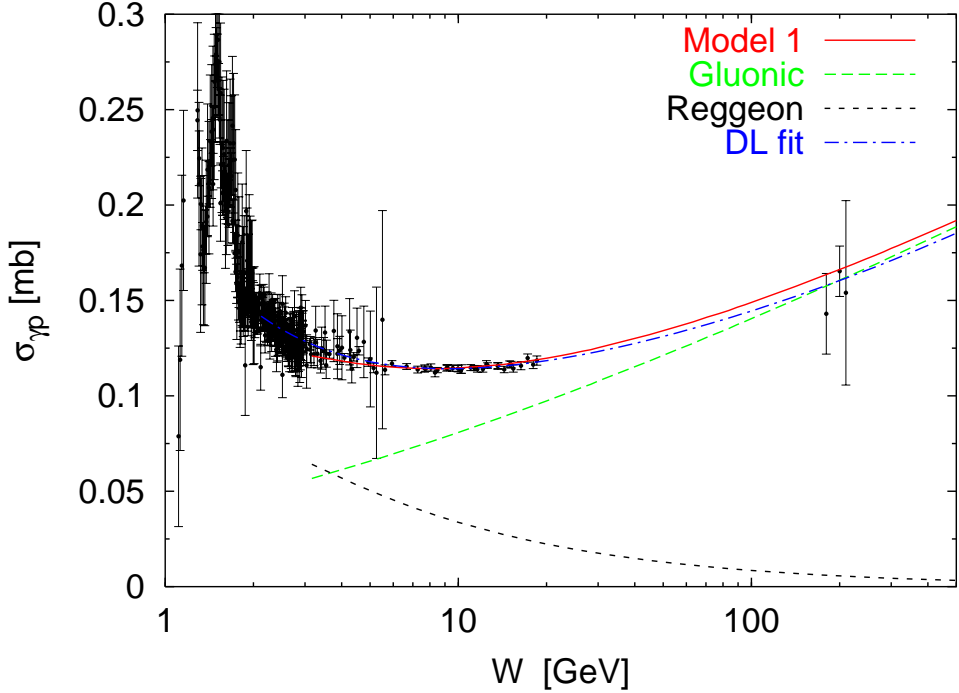}
\caption{\small\it The total $\gamma p$ cross-section -- predictions from the GBW model with
the light quark mass $m_q$ set to $0.21$~GeV and the charmed quark mass $m_c =1.3$~GeV,
supplemented by the reggeon term (\ref{regggp}), compared to data and
to the Donnachie-Landshoff fit \cite{DLgp}.
Also shown are the gluonic and reggeon components of the full result in our model.
The curves are cut at $W=3$~GeV.}

\label{realgp}
\end{center}
\end{figure}

As stated above, we modify the quark mass of the GBW saturation model 
and the Donnachie-Dosch-Rueter parametrisation for the subleading reggeon. 
Certainly, one has to ensure this change does not spoil the quality of the GBW description.
Besides that, it is necessary to check that the reggeon term with the
modified exponent $\eta=0.3$ allows the extension of the GBW model
for the $\gamma p$ total cross-section down to low values of
$W \sim 3$~GeV as well.
Thus we calculated the dipole-proton scattering contribution
using the original GBW approach, with the light quark mass, $m_q$, set
to 0.21~GeV, as in Model 1, and added the reggeon term
\be
\sigma_{\gamma p} ^R (W^2) \; = \;
A_{\gamma p}\; \left( {W^2 \over 1\,{\rm GeV}^2} \right) ^{-\eta} ,
\label{regggp}
\ee
where $A_{\gamma p}$ was fitted to data and the best value reads 
$A_{\gamma p}=0.135~{\rm mb}$.
The result is given in Fig.~\ref{realgp}, where the cross-section
from Model~1 is compared to the data, taken from Ref.~\cite{PDG}, and to the classical
Donnachie-Landshoff fit \cite{DLgp}.
In the same figure we also show the decomposition of the total
cross-section into the gluonic contribution, given by the saturation model
and the reggeon component. Both contributions have been multiplied by
a correction factor of the form
$(1-\bar x)^{2n_{\rm\tiny spect}-1}$, as described in Sec.\ 2.5, with
$n_{\rm\small spect}=2$ for the reggeon exchange, and
$n_{\rm\small spect}=4$ for the dipole-proton scattering.
The fitted curve, with only one free parameter  $A_{\gamma p}$ follows the
data accurately, suggesting that the model has certain universal properties.

\subsection{Total $\gamma\gamma$ cross-section}

\begin{figure}[t]
\begin{center}
a) \hspace{1em}
\epsfig{width= 0.5\columnwidth,file=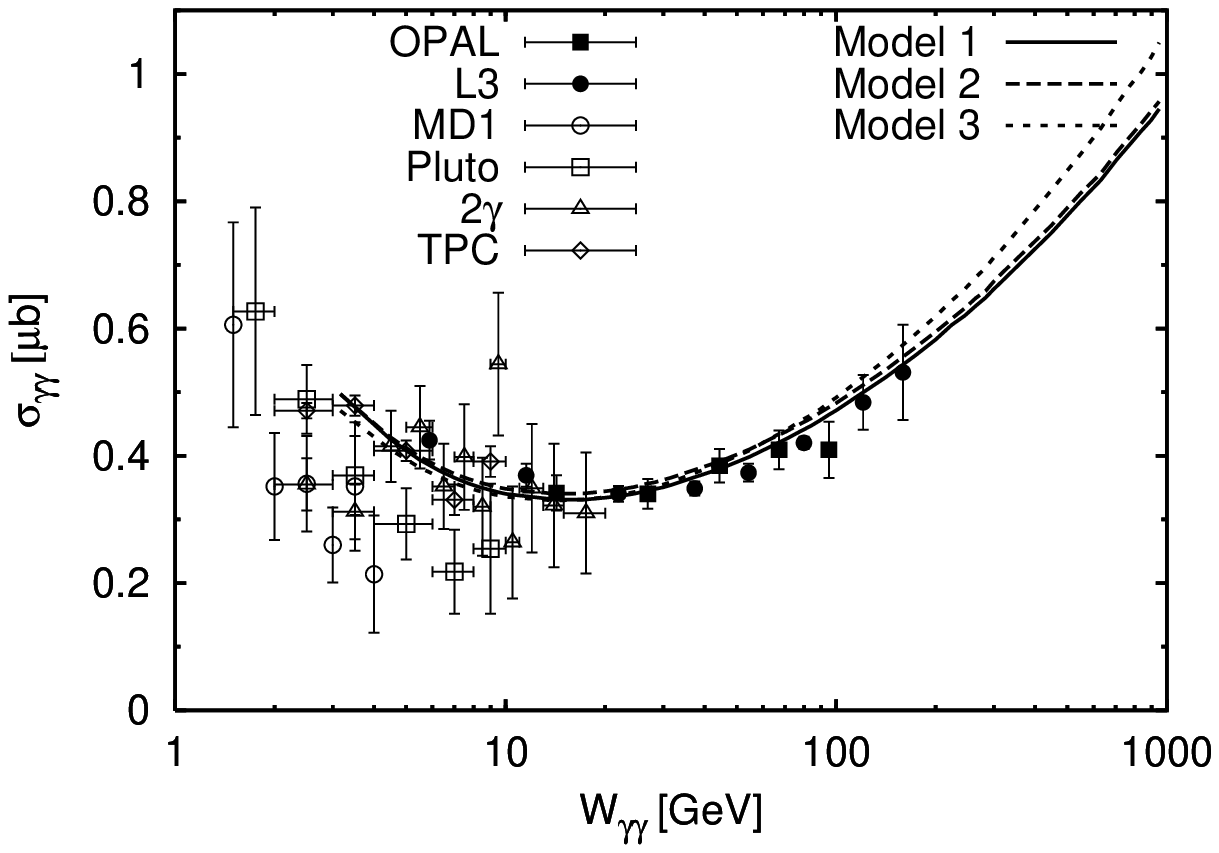} \hspace{4em} \\
b) \hspace{1em}
\epsfig{width= 0.5\columnwidth,file=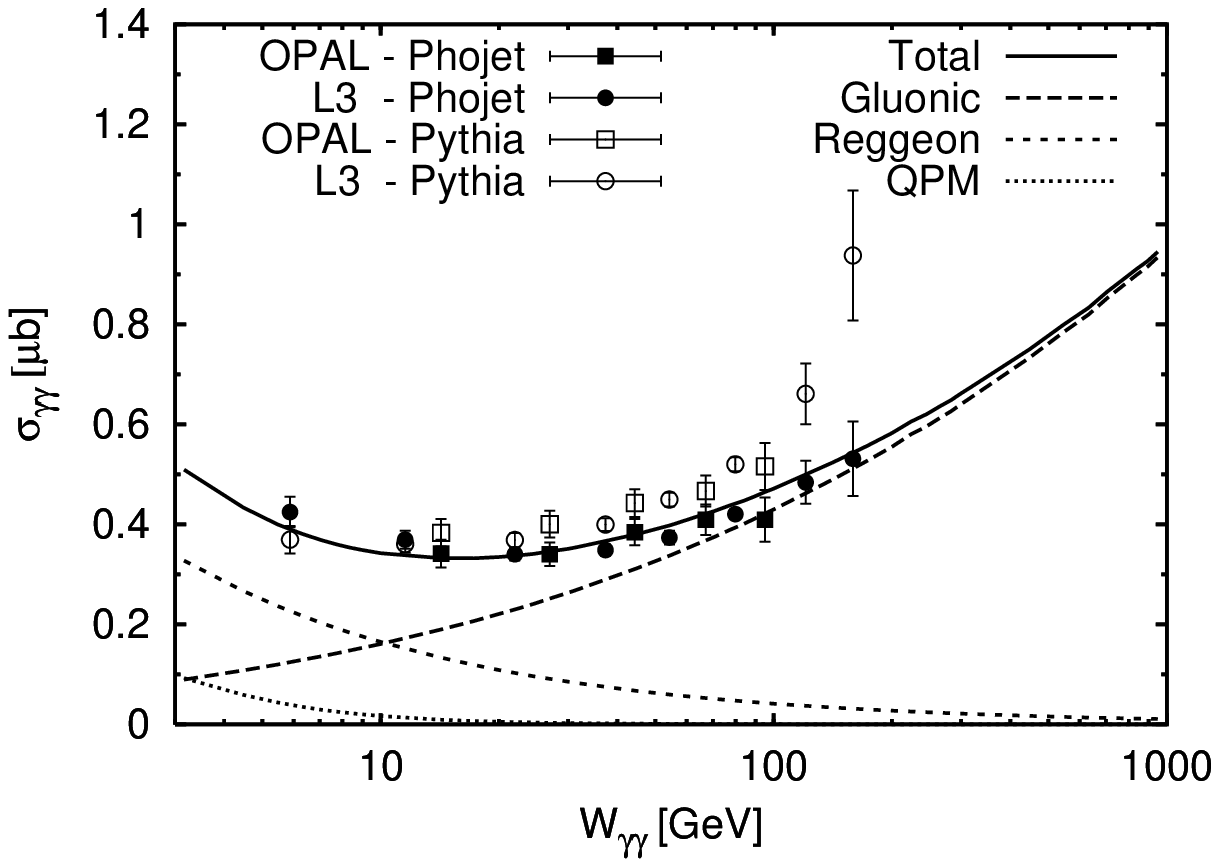} \\
\caption{\small\it The total $\gamma\gamma$ cross-section: (a) data confronted with predictions from
 all three Models and (b) contributions to the Model~1 result:
 gluonic, reggeon and QPM.}
\label{real-all}
\end{center}
\end{figure}

The available data for the $\gamma\gamma$ total cross-section range
from the $\gamma\gamma$  energy $W$ equal to about 1~GeV up to
about 160~GeV \cite{GGTOTEXP1,GGTOTEXP2}, see Fig.\ \ref{real-all}.
The experimental errors of the data are, unfortunately, rather large.
One of the reasons is that those data were taken for virtual photons
coming from electron beams and then the results were extrapolated to zero 
virtualities. 
Another problem appeared to be very important in LEP measurements where the 
incoming $e^+$ and $e^-$, and a substantial fraction of the produced hadrons 
go into the beam pipe and cannot be detected. 
Extraction of the actual $\gamma\gamma$ collision energy is
therefore needed from the visible energy, which is a model dependent
procedure and introduces large systematic errors. In particular,
it is well known, that the data for the $\gamma\gamma$ total cross-section from 
LEP depend on the Monte Carlo method applied for the unfolding.

\begin{figure}[t]
\begin{center}
\begin{tabular}{cc}
\epsfig{width= 0.45\columnwidth,file=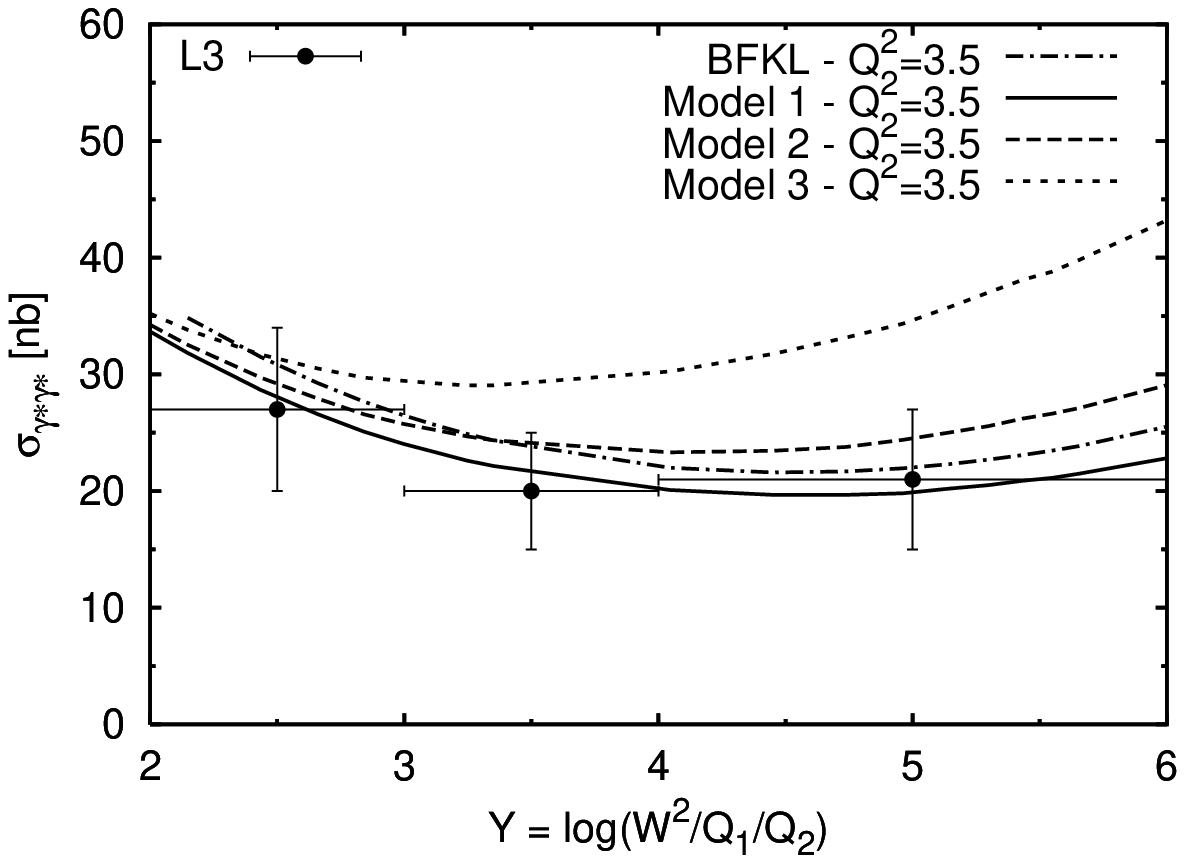} &
\epsfig{width= 0.45\columnwidth,file=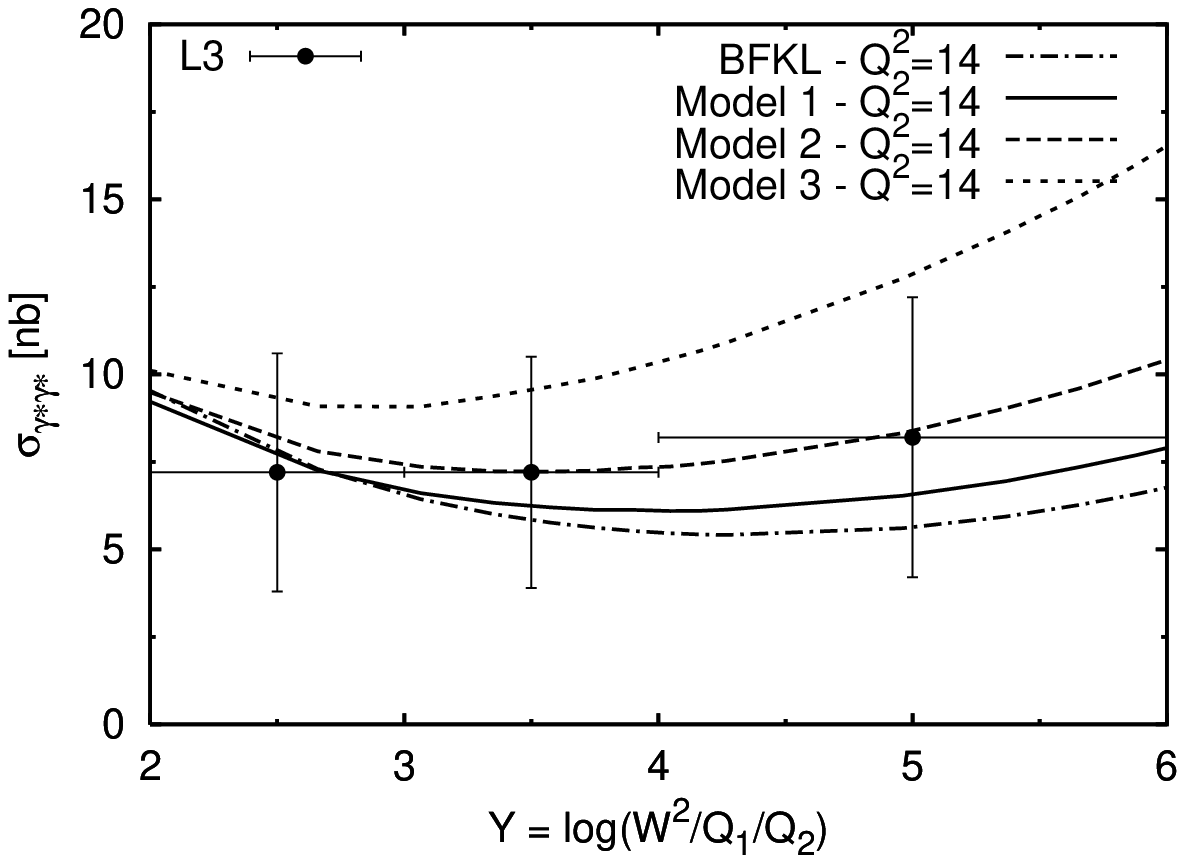}\\
a) & b) \\
\end{tabular}
\epsfig{width= 0.45\columnwidth,file=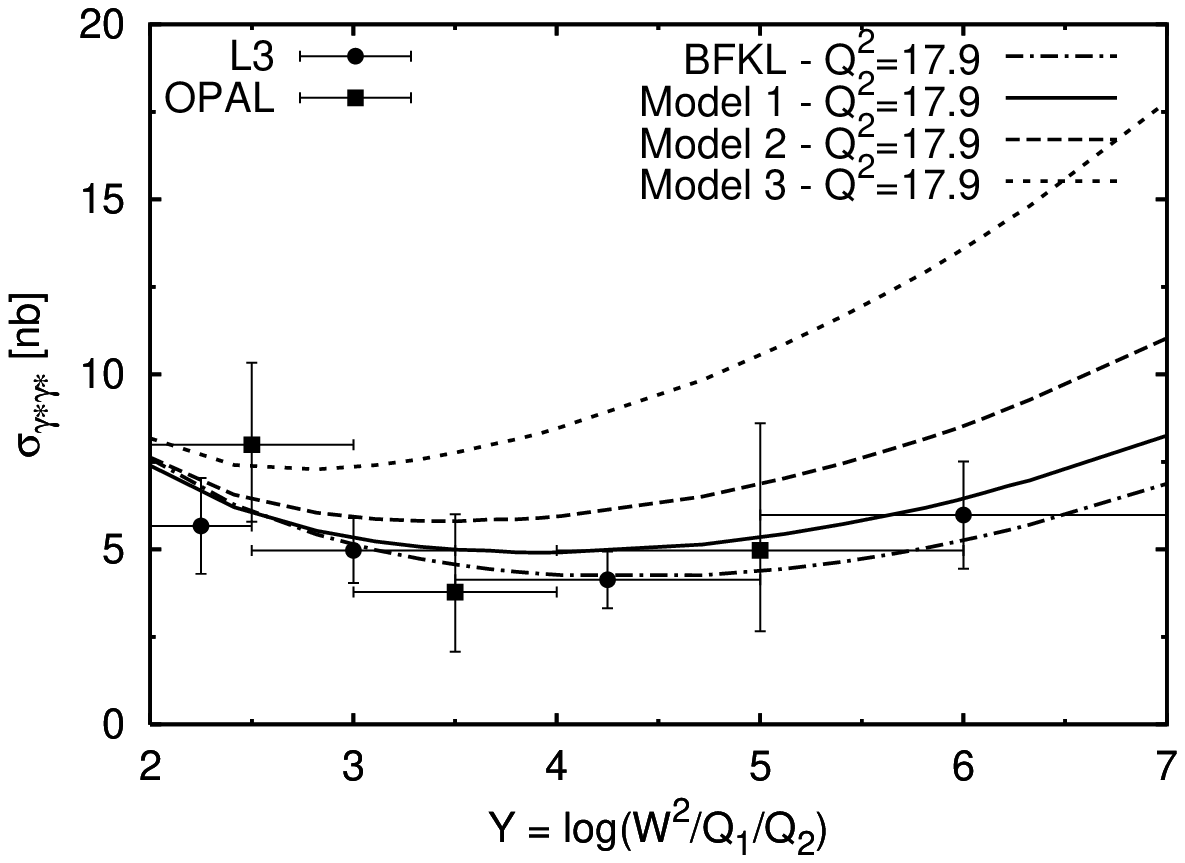} \\
c) \\
\caption{\small\it
Total $\gamma^*\gamma^*$ cross-section for
            (a) $Q ^2= 3.5~{\rm GeV}^2$,
            (b) $Q^2 = 14~{\rm GeV}^2$ and
            (c) $Q ^2=17.9~{\rm GeV}^2$ -- comparison between LEP data 
            and the Models plotted as a function of $Y= \ln (W^2/Q^2)$.
            Also shown is the result of Ref.~\cite{JKLM} based on the
            BFKL formalism with subleading corrections, supplemented by
            the QPM term, the soft pomeron and the subleading reggeon
            contributions.}
\label{virt-all}
\end{center}
\end{figure}

The data interpreted with \textsc{Pythia}\cite{pythia} tend to be larger and exhibit
a steeper rise with $W$ than those unfolded with \textsc{Phojet}\cite{phojet}
(see Fig.~\ref{real-all}b).
In the saturation model it is difficult to obtain a cross-section
consistent with the \textsc{Pythia}-unfolded data. Besides that, \textsc{Phojet} is a 
Monte Carlo program dedicated to describe two-photon interactions and the description of 
the crucial hadron emissions close to two-photon collision axis, 
is elaborated in more detail. Thus we choose to follow the
\textsc{Phojet} unfolded data in our analysis. In a more conservative approach
one should include the difference between the cross-sections unfolded with
different programs into the systematic error. This would only make the data
less constraining and would not spoil the quality of the fit.

In Fig.~\ref{real-all}a we show the total $\gamma\gamma$ cross-section 
from the Models, obtained using eq.\ (\ref{total}) with $i=j=T$.
The agreement with data is very good down to $W \simeq 3$~GeV.
It is interesting to observe, that the Models strongly favour
the \textsc{Phojet} unfolded data, and that the energy dependence of the total 
$\gamma\gamma$ cross-section (\textsc{Phojet}-unfolded) at high~$W$ is very 
well reproduced by all three Models, see Fig.\ \ref{real-all}b.
Recall, that the steeper $W$-dependence found
in the two-photon cross-section, as compared to the hadronic
and the photoproduction cross-sections, is naturally explained by
the presence of additional factors of $\ln^2 W$ and $\ln\, W$
respectively, as discussed in Section 2.3. Predictions for $W$ in the
range to be probed in future linear colliders (i.e. $W<1$~TeV) are
stable against variations of the details of the saturation models, provided that
the models are adjusted to fit the available data.

\subsection{Total $\gamma^*\gamma^*$ cross-section}

The data \cite{DTL3,DTOPAL} for the total $\gamma^*\gamma^*$ cross-section are extracted
from so-called double-tagged events, that is from $e^+e^-$ events
in which both the scattered electrons are measured and hadrons are produced.
In such events  measurement of the kinematical variables of the leptons
determines both the virtualities $Q_1^2$ and $Q_2^2$ of the colliding photons and
the collision energy $W$. The tagging angles in LEP experiments restrict the
virtualities to be similar, i.e $Q_1^2 \sim Q_2^2 = Q^2$.
The data are available from LEP for average values
$Q ^2=3.5~{\rm GeV}^2$, $14~{\rm GeV}^2$ and $Q ^2=17.9~{\rm GeV}^2$
in a wide range of $W$.

In Fig.\ \ref{virt-all}a,b,c those data are compared with the curves from 
the Models.
As an estimate of the total $\gamma^*\gamma^*$ cross-section 
we use a simple sum of the cross-sections $\sigma_{ij} ^{\rm\small tot}$ 
(eq.\ (\ref{total})) over transverse and longitudinal polarisations $i$ and~$j$
of both photons. 
In addition we plot also the prediction obtained in Ref.~\cite{JKLM} by solving
the BFKL equation with non-leading effects, and added phenomenological
soft pomeron and reggeon contributions and the QPM term.
The latter prediction was found to describe the measured $e^+ e^-$ differential
cross-section for hadron production in double tagged events \cite{JKLM}.
As can be seen, Model~1 fits data as well as the result based on the BFKL solution.
Model~2 is slightly worse than Model~1 and Model~3 may be rejected.

\begin{figure}[t]
\begin{center}
\begin{tabular}{cc}
\epsfig{width= 0.45\columnwidth,file=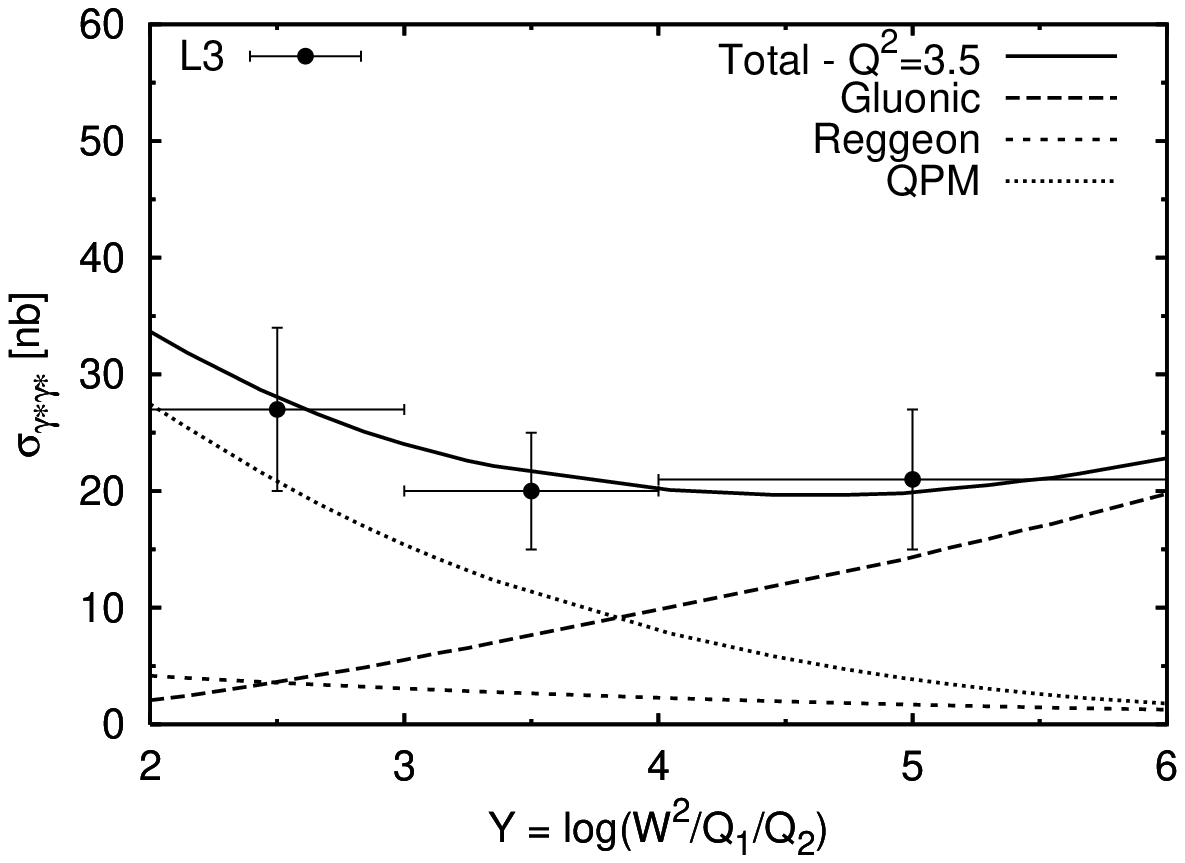} &
\epsfig{width= 0.45\columnwidth,file=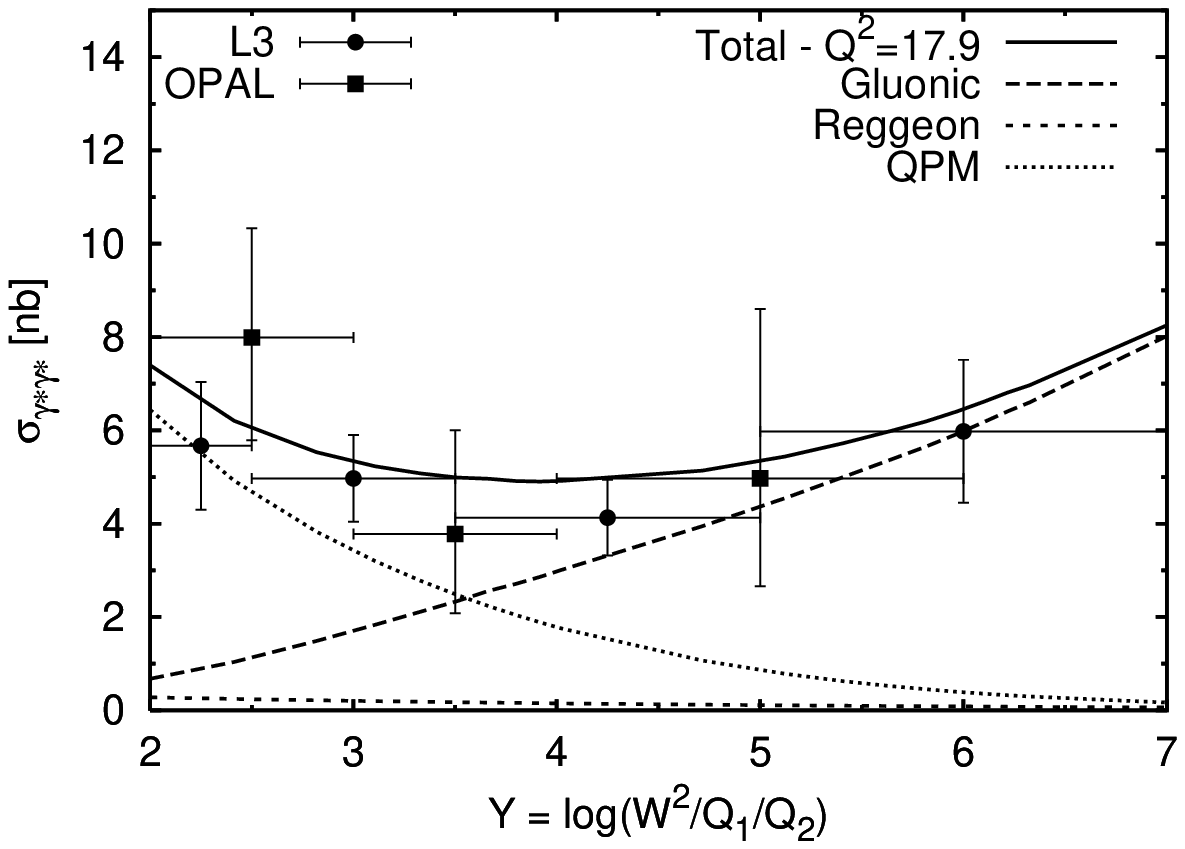} \\
a) & b) \\
\end{tabular}
\caption{\small\it The total $\gamma^*\gamma^*$ cross-section -- the decomposition
of Model 1 into QPM, gluonic and reggeon components for (a) low virtuality
$Q^2 = 3.5~{\rm GeV}^2$ and (b) large virtuality
$Q^2 = 17.9~{\rm GeV}^2$.}
\label{virt-separated}
\end{center}
\end{figure}

Since the virtuality $Q^2$ is high, the unitarity corrections are not
important here. For the same reason, the results are not sensitive to the
choice of the quark masses and the parameters of the reggeon term.
As seen in Fig.~\ref{virt-separated}, where the components
of the $\gamma^*\gamma^*$ total cross-section from Model~1 are 
plotted, the cross-section is dominated by the QPM and the `pomeron' contributions.
Moreover, the perturbative approximation for the photon wave function is
fully justified in this case. Thus, in this measurement the form of the dipole-dipole
cross-section is directly probed.

\begin{figure}[t]
\begin{center}
\begin{tabular}{cc}
\epsfig{width= 0.45\columnwidth,file=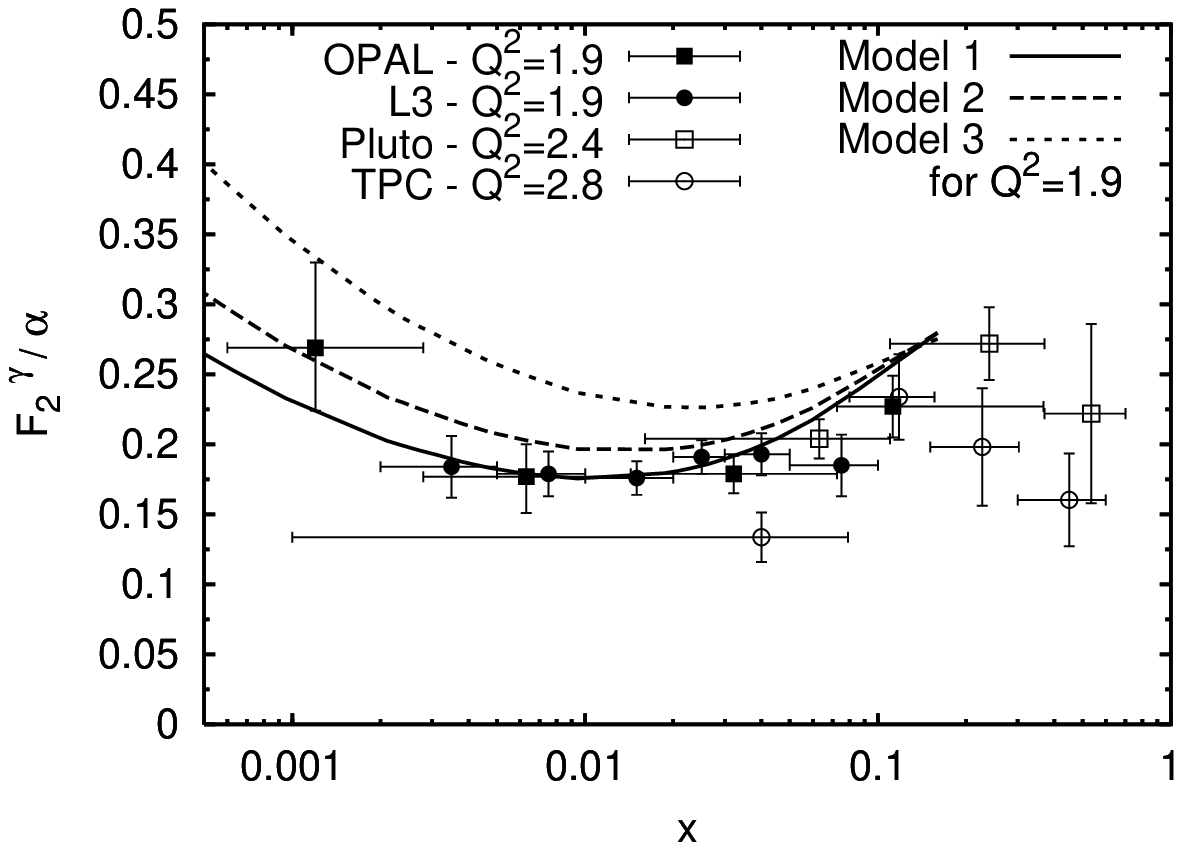} &
\epsfig{width= 0.45\columnwidth,file=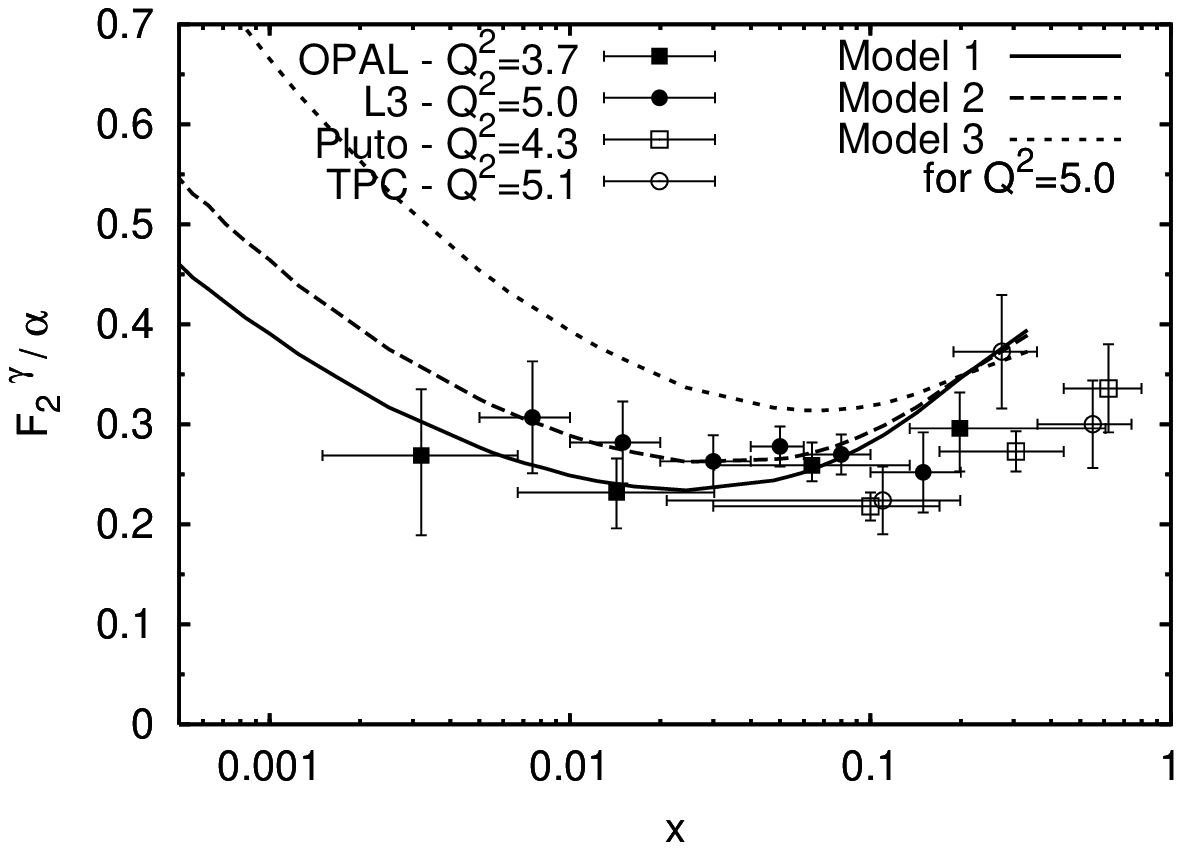}\\
a) & b) \\
\epsfig{width= 0.45\columnwidth,file=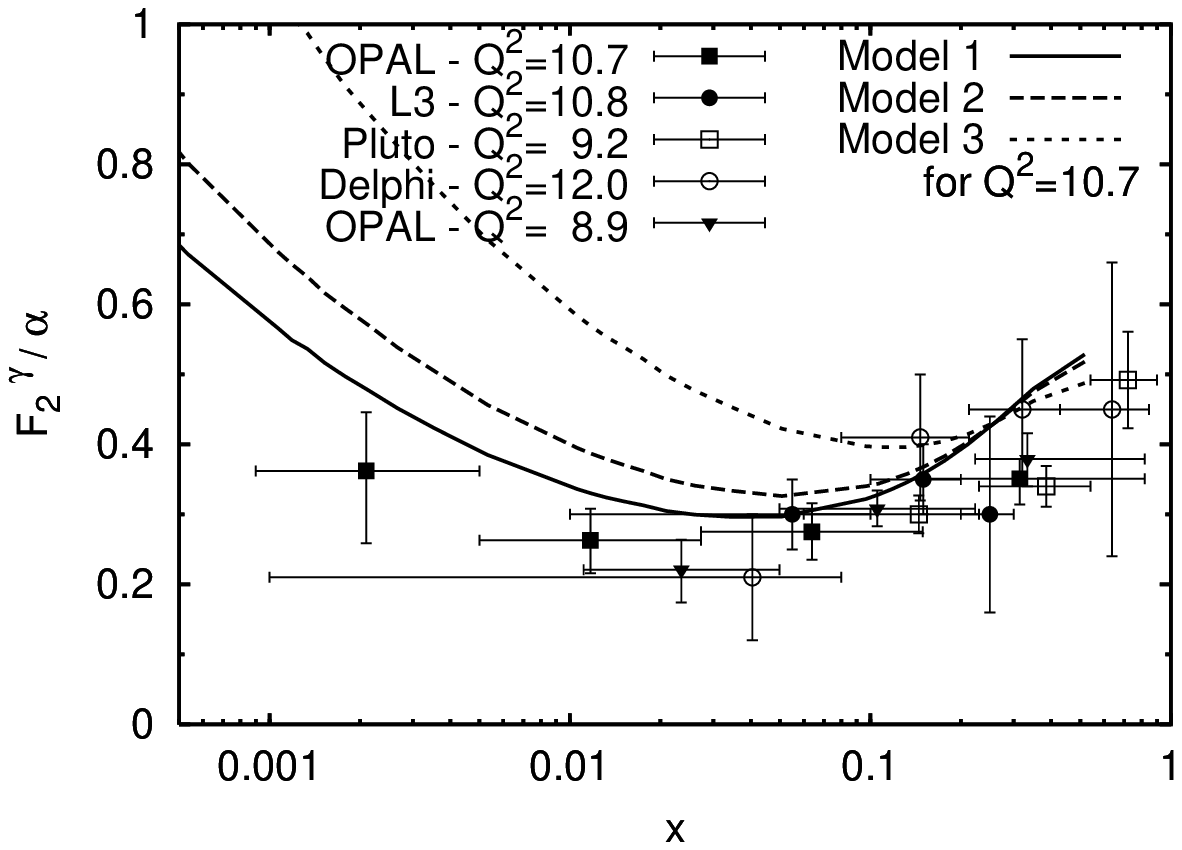} &
\epsfig{width= 0.45\columnwidth,file=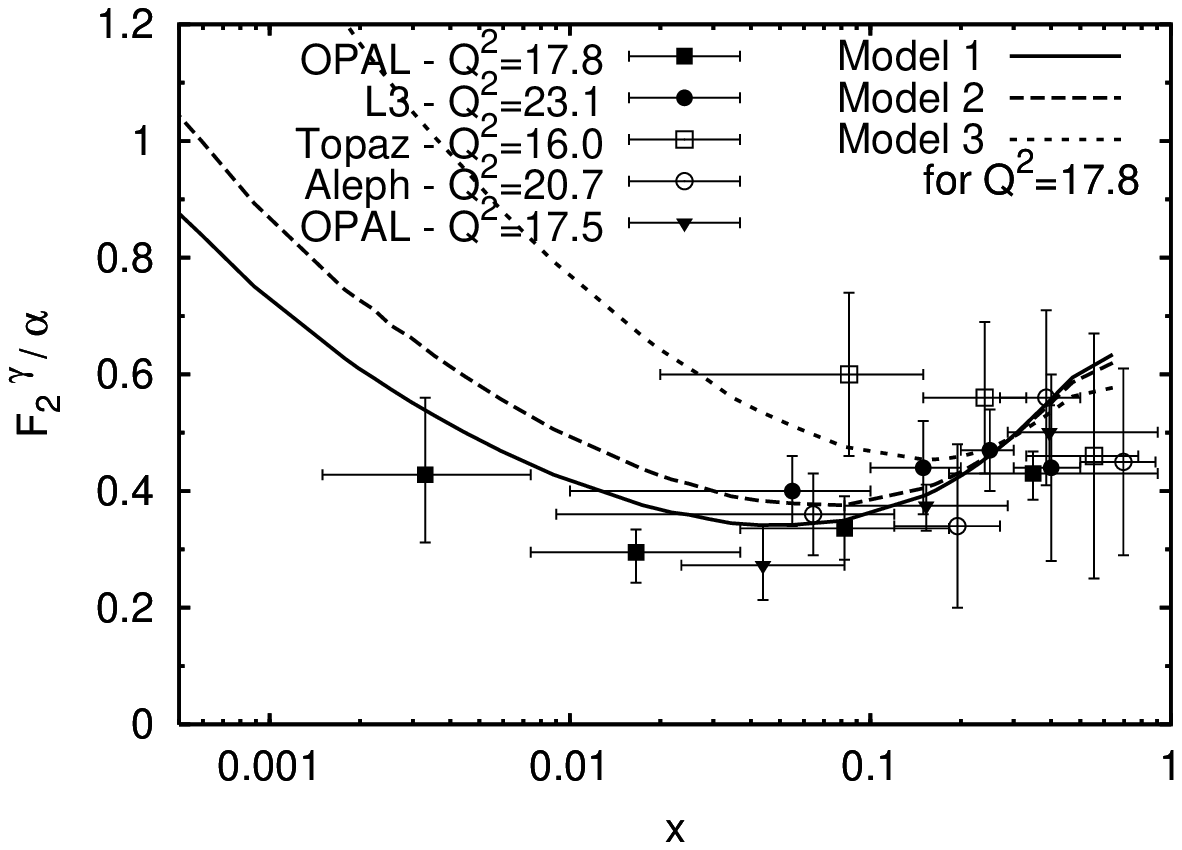} \\
c) & d) \\
\end{tabular}
\caption{\small\it The photon structure function $F_2 ^{\gamma}(x,Q^2)$:
the experimental data compared to predictions following from the
Models for various $Q^2$: (a) from 1.9 to 2.8~GeV$^2$, (b) from 3.7 to 5.1~GeV$^2$,
(c) from 8.9 to 12.0~GeV$^2$ and (d) from 16.0 to 23.1~GeV$^2$.
The curves are cut at values of $x$ corresponding to $W=3$~GeV.}
\label{f2-all}
\end{center}
\end{figure}

\subsection{Photon structure}

The quasi-real photon structure may be probed in single tagged $e^+e^-$ events.
In this case one of the electrons scatters with a larger momentum transfer $Q_1^2=Q^2$
which corresponds to the emitted photon virtuality and the other electron
scatters at a low angle, producing predominantly a virtual photon with very
low virtuality $Q_2 ^2\simeq 0$. Thus, the measurement of the cross-section for
the $\gamma^*(Q_1^2) \gamma^* (Q_2^2)$ collision at the energy $W$ can be used
to  extract the almost real photon  structure function $F_2^{\gamma}(x,Q^2)$,
with $x=Q^2/(W^2+Q^2)$, see eq.~(\ref{fgg}).

Essentially, the parameters of the models are constrained by the data
for the total $\gamma\gamma$ and  $\gamma^*\gamma^*$ cross-sections so
here we are presenting a parameter free result.
In Fig.~\ref{f2-all}  we show  the comparison of our predictions with the experimental
data \cite{F2EXP1,F2EXP2}
 for the  virtuality $Q^2$  in the range from
(a)~1.9 to~2.8~GeV$^2$,
(b)~3.7 to~5.1~GeV$^2$,
(c)~8.9 to~12.0~GeV$^2$
and finally (d)~from 16.0 to~23.1~GeV$^2$.
Note, that in each plot the data for various virtualities are
combined, which may give rise to systematic effects, see for instance
Fig.~\ref{f2-all}b. In each plot the value of virtuality $Q^2$ adopted to obtain
the theoretical curve is indicated and was selected to match the average 
value $Q^2$ of the data-set containing the best data at low $x$. 
We also show in Fig.~\ref{f2-components}
the contributions to $F_2 ^{\gamma}$ in Model~1.
As already stated, the importance of the reggeon
term is not very large and decreases with increasing $Q^2$.

\begin{figure}[t]
\begin{center}
\begin{tabular}{cc}
\epsfig{width= 0.45\columnwidth,file=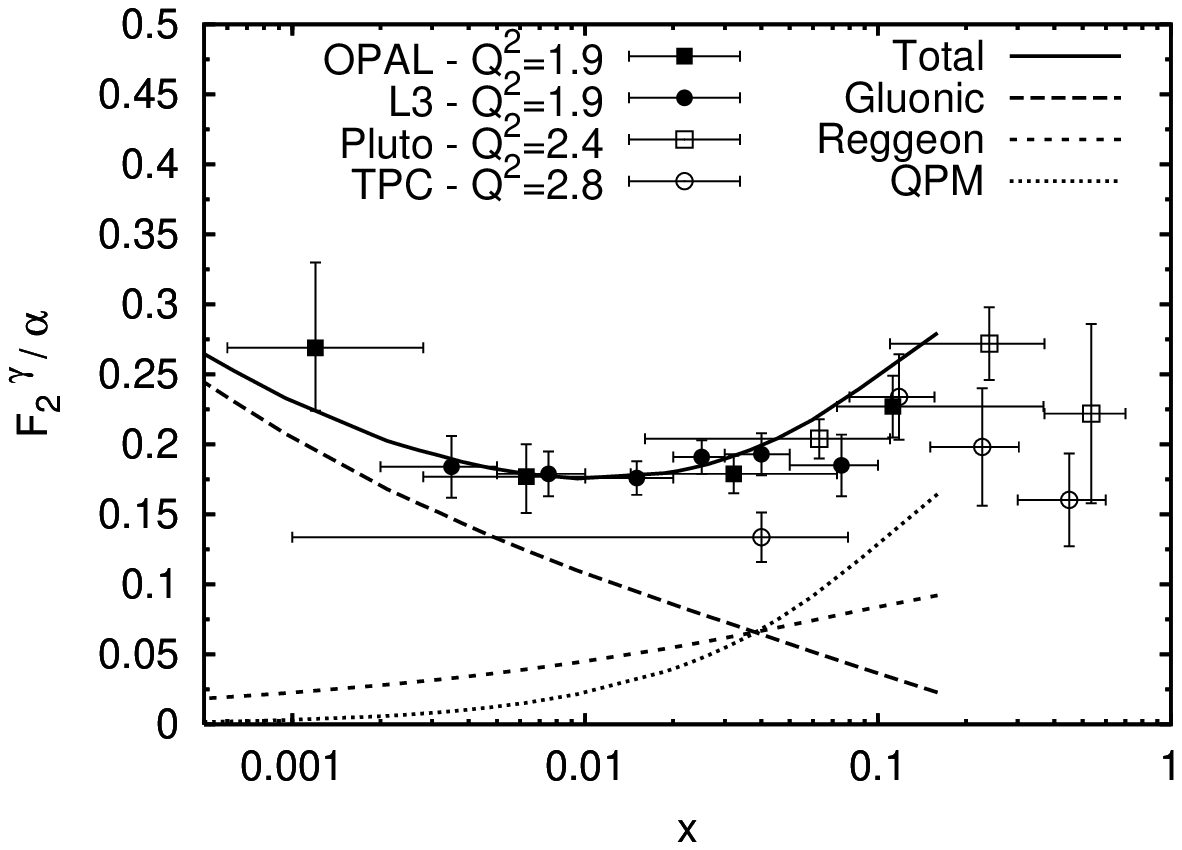} &
\epsfig{width= 0.45\columnwidth,file=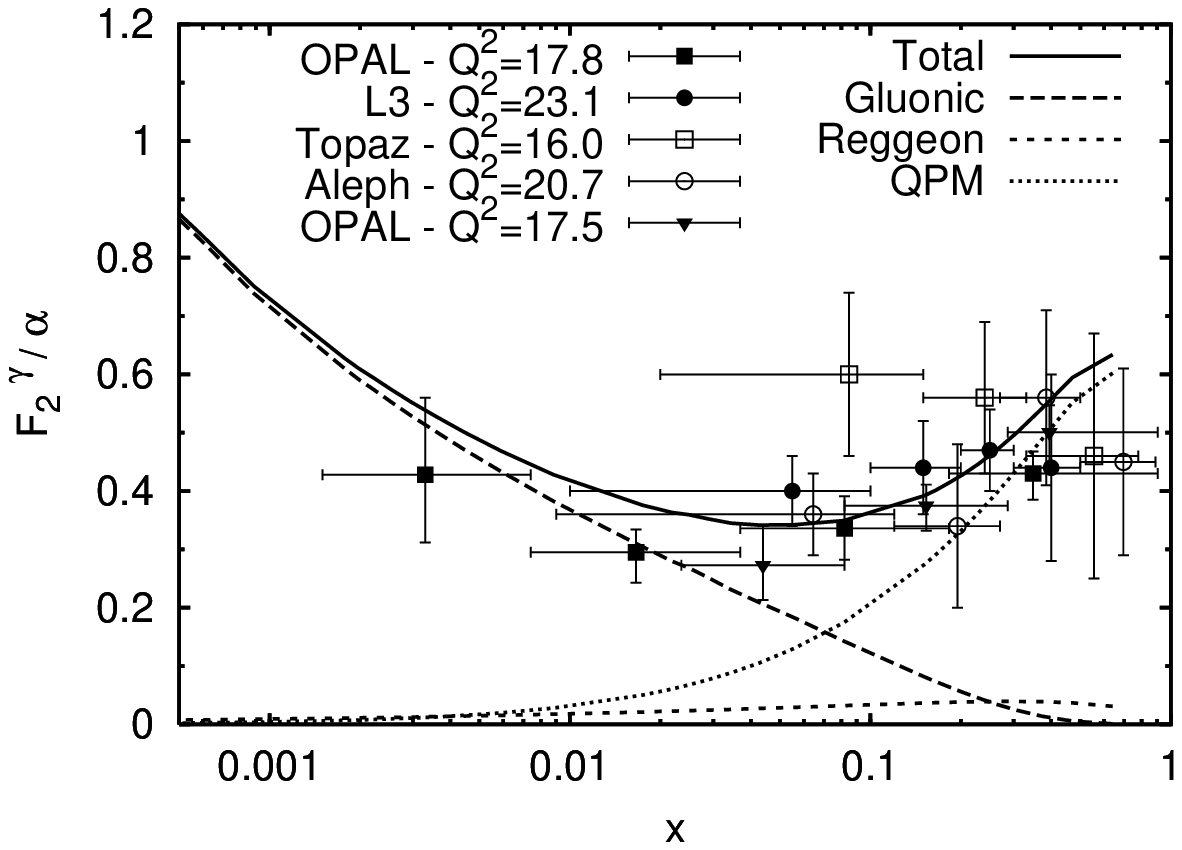} \\
a) & b) \\
\end{tabular}
\caption{\small\it The photon structure function $F_2 ^{\gamma}(x,Q^2)$ from Model~1
compared to the experimental data for (a) low $Q^2$, 1.9 to 2.8~GeV$^2$,
and (b) large $Q^2$, from 16.0 to 23.1~GeV$^2$.}
\label{f2-components}
\end{center}
\end{figure}

Model 1, favoured by the $\gamma^*\gamma^*$ data provides the best description
of $F_2 ^{\gamma}$ as well. In the region of $x>0.1$
the agreement of the Model~1 with data is surprisingly good
which was not {\it a~priori} expected from the model based on the large
energy approximation.
Note however, that a systematic tendency occurs for all the Models 
to overestimate the data for larger $Q^2$.

It is straightforward to obtain in this framework  predictions for
the virtual photon (with $Q_2^2 = P^2$) structure function $F_2^{\gamma^*}(x,Q^2;P^2)$
in the low $x$ domain. However, there exist only very few data on this observable
so we do not present our predictions for this quantity.
Nevertheless, possible experimental study of $F_2^{\gamma^*}(x,Q^2;P^2)$
would certainly provide  another interesting test of the saturation model.

\subsection{Heavy flavour production}

Another interesting process which we have studied in the dipole model
is the production of heavy flavours (charm and bottom) in $\gamma\gamma$
collisions. Heavy quarks can be produced by three mechanisms:
\begin{enumerate}
\item The direct production in which both photons couple to the same
      heavy quark line, which corresponds to a component of the
      quark box diagram (see Fig.~\ref{charmdiag}a).
\item The direct photoproduction off the resolved photon, which would involve
      a fluctuation of one of the photons into a heavy quark-antiquark pair
      and scattering of the pair off the other, resolved photon by exchange of
      gluons. This phenomenon is accounted for in the dipole model and the
      cross-section may be obtained by restricting the sum over the
      flavours in eq.~(\ref{psisum}) to the case, in which at least one dipole
      is composed of the heavy quarks only (see Fig.~\ref{charmdiag}b).
\item  The hard fragmentation and rescattering contributions.
       The first one corresponds to production of a heavy quark pair
       in the fragmentation process of an initial light quark pair.
       The initial pairs can be produced either through the box
       diagram (Fig.~\ref{charmdiag}c) or as the colour dipoles
       (Fig.~\ref{charmdiag}d) which are present in the model.
       Note also that in the saturation model, in the case of real photons
       and the original dipoles composed of light quarks,
       abundant rescattering of cascading gluons occur in which
       heavy quarks may be produced. This would be the other,
       rescattering mechanism. The estimate of such effects is, so far,
       beyond the reach of our model and we do not take into account these
       contributions.
\end{enumerate}

\begin{figure}[t]
\begin{center}
\begin{tabular}{cc}
\epsfig{width= 0.25\columnwidth,file=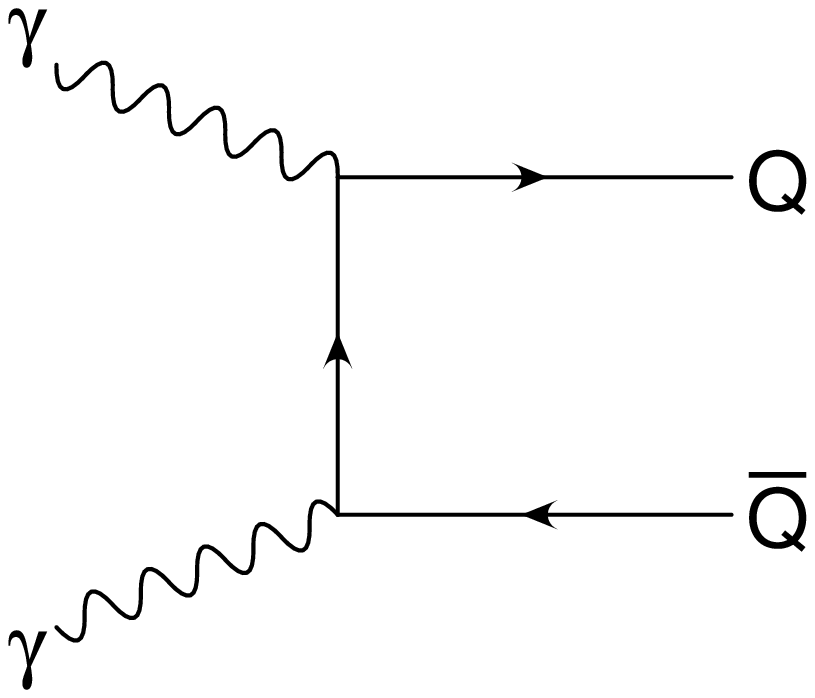} \hspace*{10mm} &
\epsfig{width= 0.25\columnwidth,file=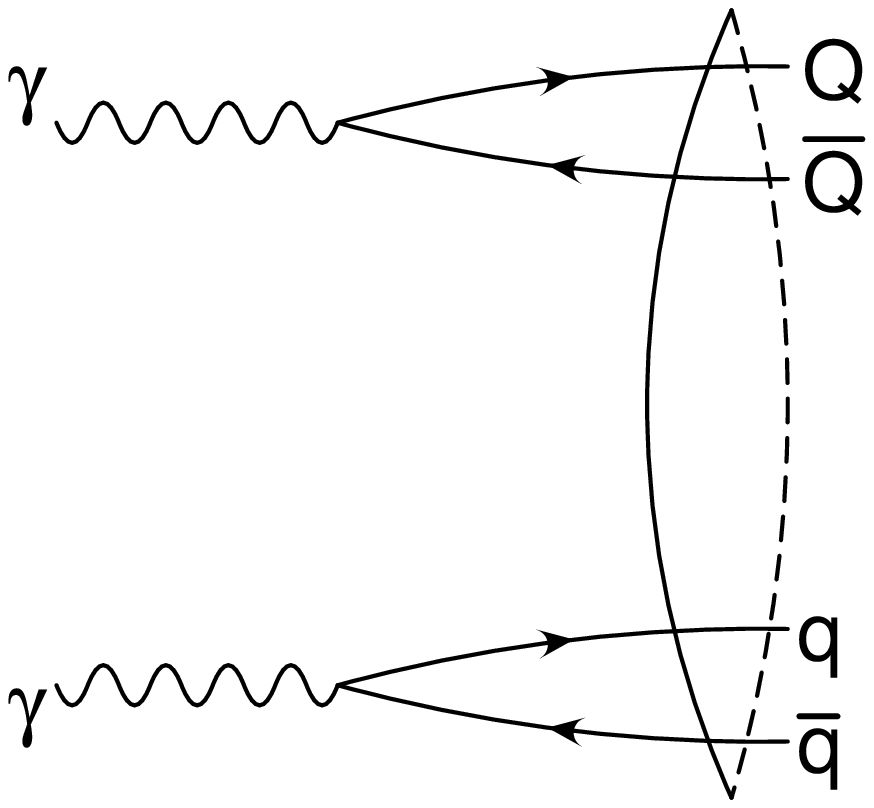}\\
a) & b) \\
\\
\epsfig{width= 0.25\columnwidth,file=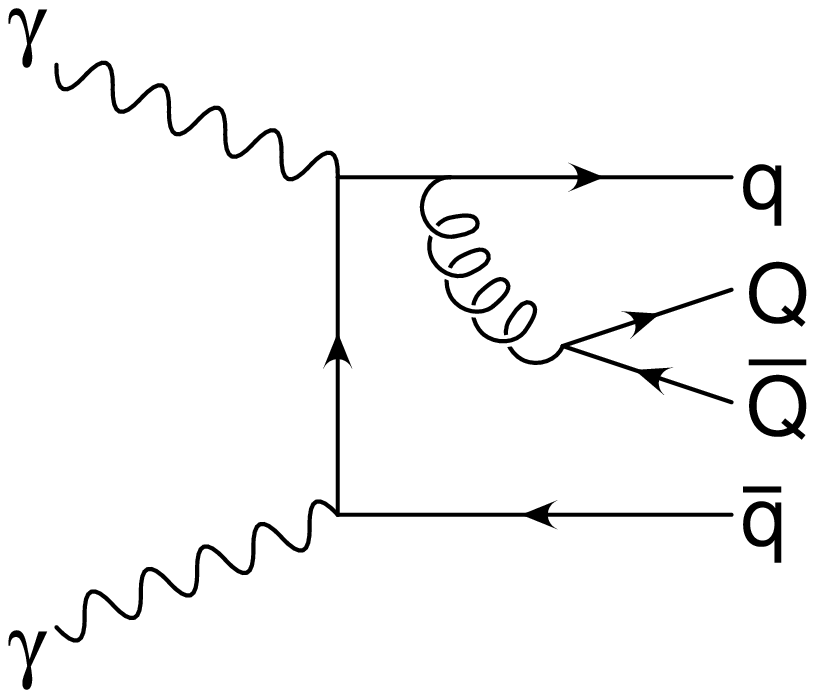} \hspace*{10mm} &
\epsfig{width= 0.25\columnwidth,file=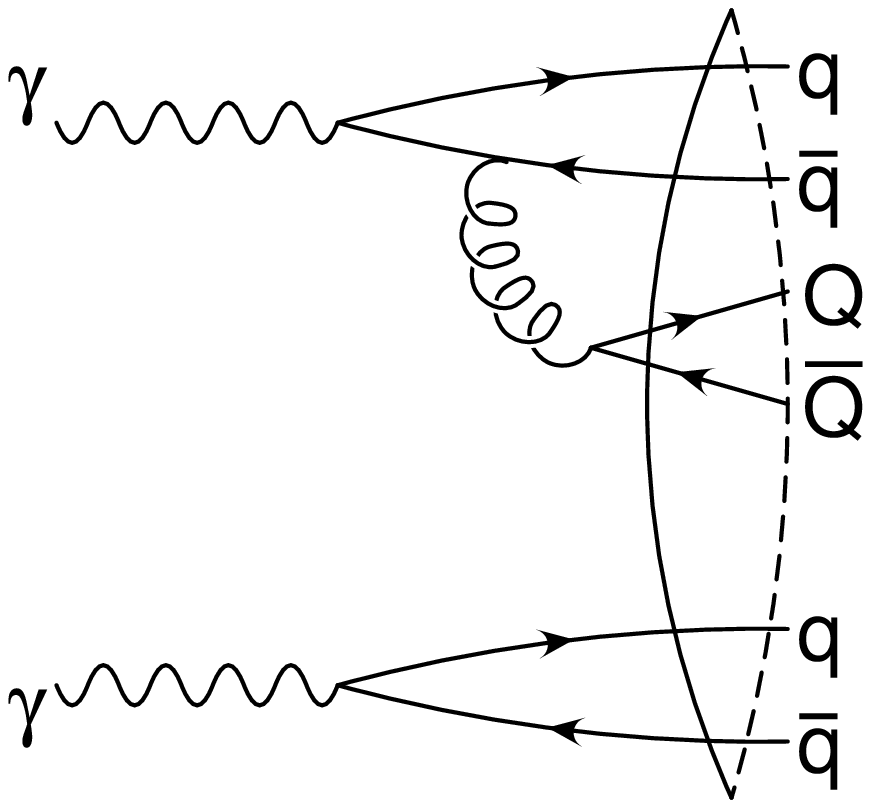}\\
c) & d) \\
\end{tabular}
\caption{\small\it Diagrams illustrating contributions to the heavy quark production,
at the amplitude level: (a) heavy quark box diagram, (b) direct production of heavy flavours in one of the dipoles, (c) example of production
of a heavy quark pair through a hard fragmentation process in a box diagram, and (d)
representation of production by hard fragmentation from a light quark dipole, or
by gluon rescattering.}
\label{charmdiag}
\end{center}
\end{figure}

The reggeon exchange is a non-perturbative phenomenon and should not
contribute to heavy flavour production, so it is assumed to vanish here.
In Fig.~\ref{charm} we plot the predictions from all three Models compared with
L3 data on charm production \cite{L3cc}. The best model, Model~1, is below the
data leaving some room  for a possible contribution from the fragmentation.
The shape of the cross-section is well reproduced.

\begin{figure}[t]
\begin{center}
\begin{tabular}{cc}
\epsfig{width= 0.45\columnwidth,file=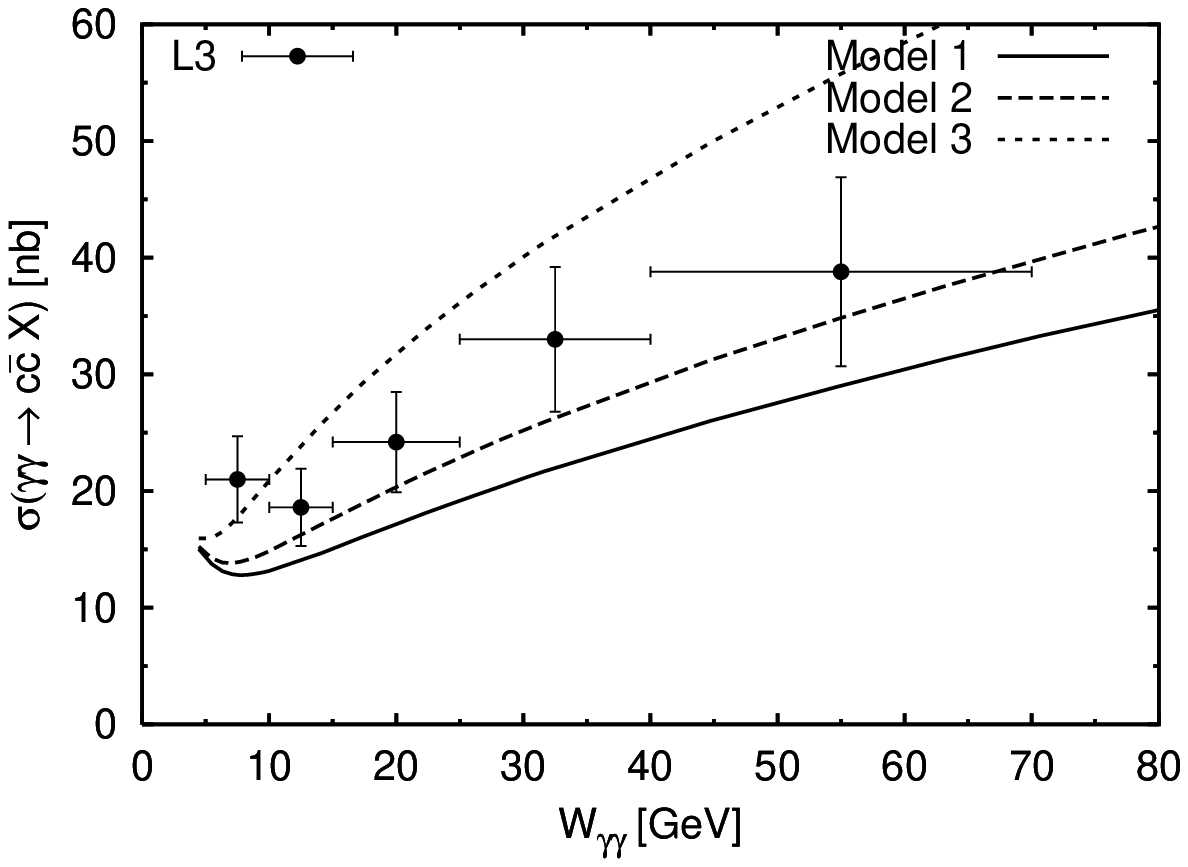} &
\epsfig{width= 0.45\columnwidth,file=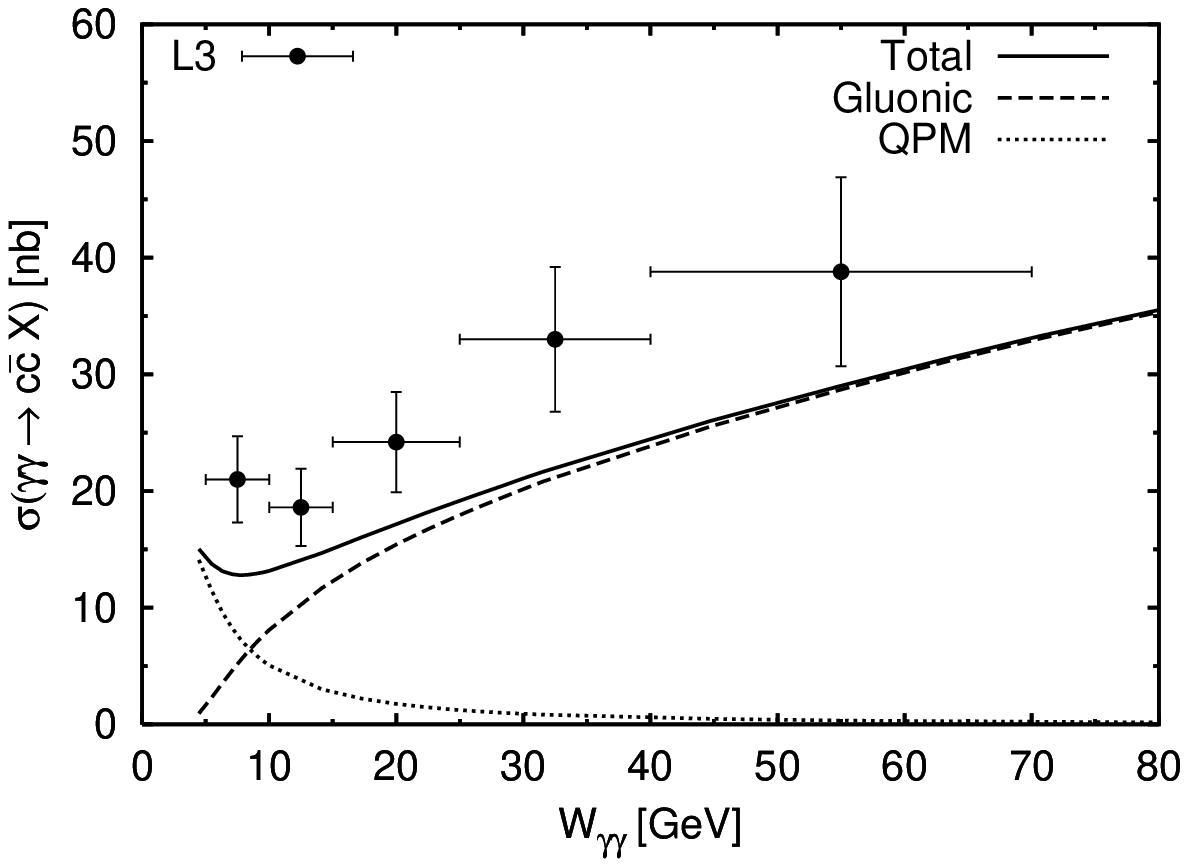} \\
a) & b) \\
\end{tabular}
\caption{\small\it The cross section for the inclusive charm production in $\gamma\gamma$
collisions: (a) results for all three Models and (b) the decomposition of
the result from Model~1 on the QPM and gluonic component.}
\label{charm}
\end{center}
\end{figure}

Production of bottom quarks in two almost real photon collisions was
investigated experimentally by the L3 \cite{L3bb} the OPAL \cite{OPbb}
collaborations. There, the measured process was
$e^+e^- \to e^+e^- b\bar{b} X$, with anti-tagged electrons
at $e^+e^-$ invariant collision energies $\sqrt{s}_{ee}$ between 189~GeV and 202~GeV.
The total cross-section for this reaction was found to be
$13.1 \pm 2.0 \, {\rm (stat)} \pm 2.4 \,{\rm (syst)}$~pb (L3) and
$14.2 \pm 2.5 \, {\rm (stat)} \pm 5 \,{\rm (syst)}$~pb (OPAL)
whereas the theoretical estimate from Model~1  for $\sqrt{s}_{ee} = 200$~GeV
gives about 5.5~pb with less than 10\% uncertainty related to the choice of $b$-quark mass.
This is significantly below the experimental data but above
the expectations of $3 \pm 1$~pb (see e.g.\ \cite{OPbb}),
based on standard QCD calculations with the use of the resolved photon
approximation.

In conclusion, the saturation model underestimates the cross-section
for production of heavy quarks and the discrepancy increases with
increasing quark mass, or perhaps, decreasing electric charge.
This may be a hint that the fragmentation and rescattering
mechanisms of heavy quark production are, indeed, important.

\section{Predictions for future colliders}

Two-photon processes  will be important at
possible future $e^+ e^-$ colliders, like TESLA, where
the available photon-photon collision energy might reach
500~GeV or even 1~TeV~\cite{TESLA}.
Thus, we give  predictions from  Model 1 for the energy dependence of
$\,\gamma^*(Q ^2) \gamma^*(Q^2)\,$, the {\em gluonic} component
$\,\tilde\sigma^G _{\gamma^*\gamma^*}(W^2,Q^2,Q^2)\,$ 
(defined by eqs.\ (\ref{master}) and (\ref{sigmaddth}))
of the total cross-section 
$\,\sigma^{\rm\small tot} _{\gamma^*\gamma^*}(Q_1^2,Q_2^2,W)\,$
for $Q_1^2 \,=\, Q_2^2 \,=\, Q^2$.
In Fig.~\ref{future-gg}a,b  we show the results in terms of a 
re-scaled quantity 
$\,\bar{Q} ^2 \tilde\sigma^G _{\gamma^*\gamma^*}(W^2,Q^2,Q^2)\,$, with
$\,\bar{Q}^2 = \max(Q^2,4m_q^2)\,$
for various $Q^2$ between 0~and 10~GeV$^2$ (Fig.~\ref{future-gg}a)
and between 10 and 200~GeV$^2$ (Fig.~\ref{future-gg}b).
Of course, the cross-section for gluon exchange
$\,\tilde\sigma^G_{\gamma^*\gamma^*}(W^2,Q^2,Q^2)\,$ has to be
combined with the standard QPM and reggeon terms in order to get
a complete description of the total cross-section, as described
in Sec.\ 2.6.

\begin{figure}[t]
\begin{center}
\begin{tabular}{cc}
\epsfig{width= 0.45\columnwidth,file=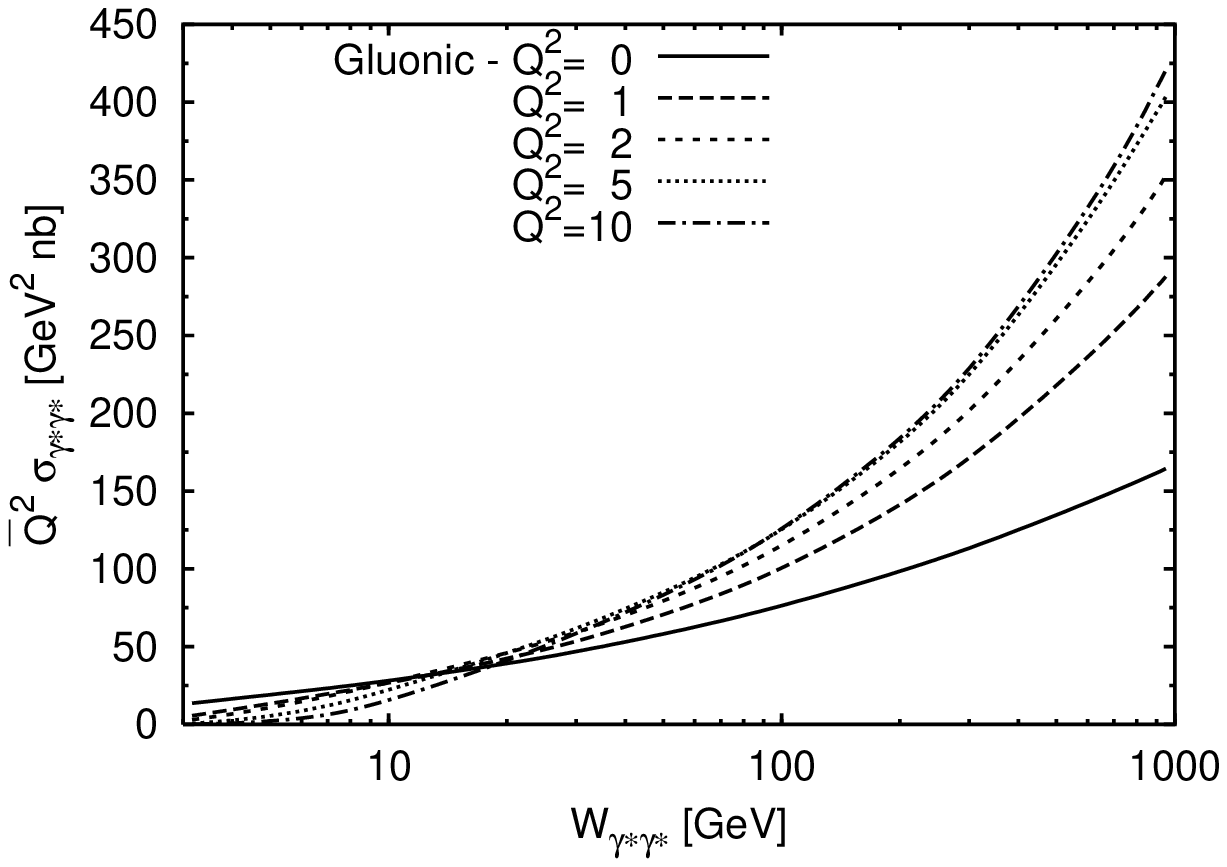} &
\epsfig{width= 0.45\columnwidth,file=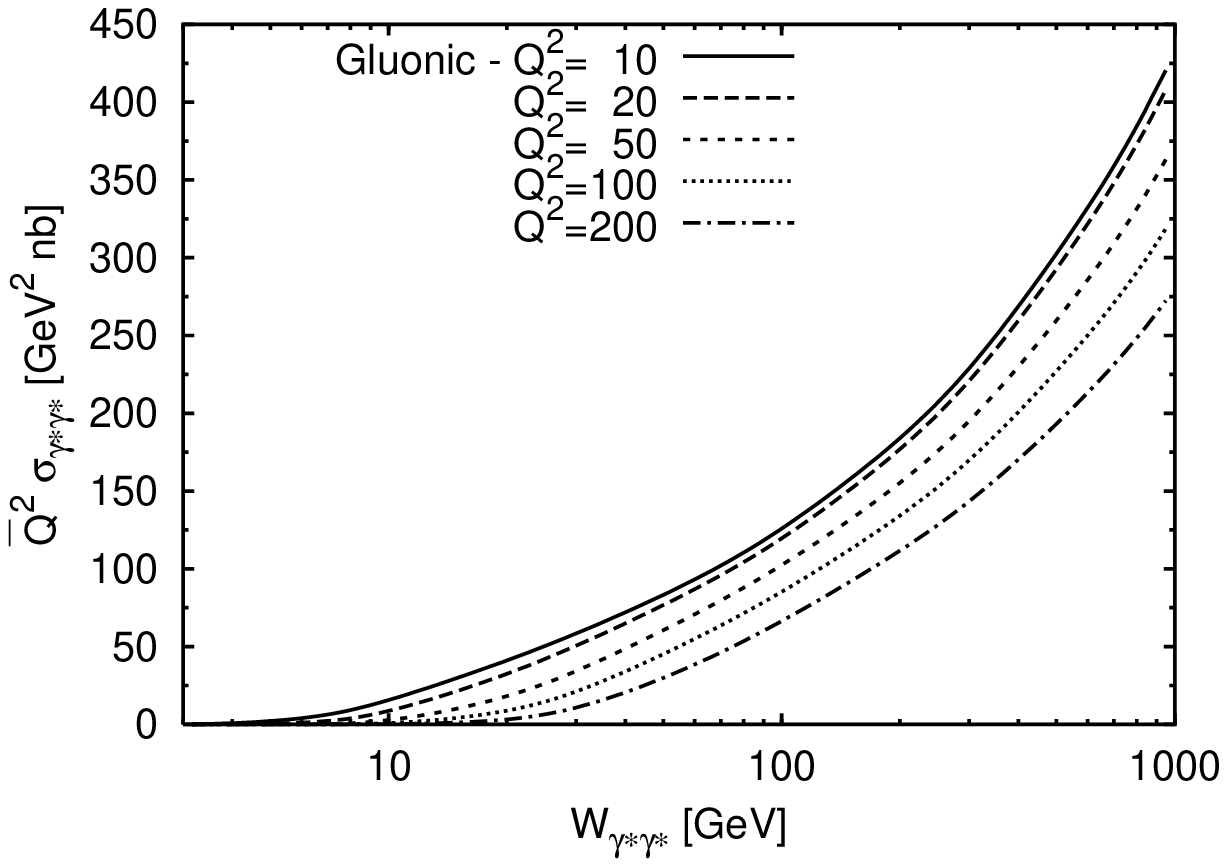} \\
a) & b) \\
\end{tabular}
\epsfig{width= 0.45\columnwidth,file=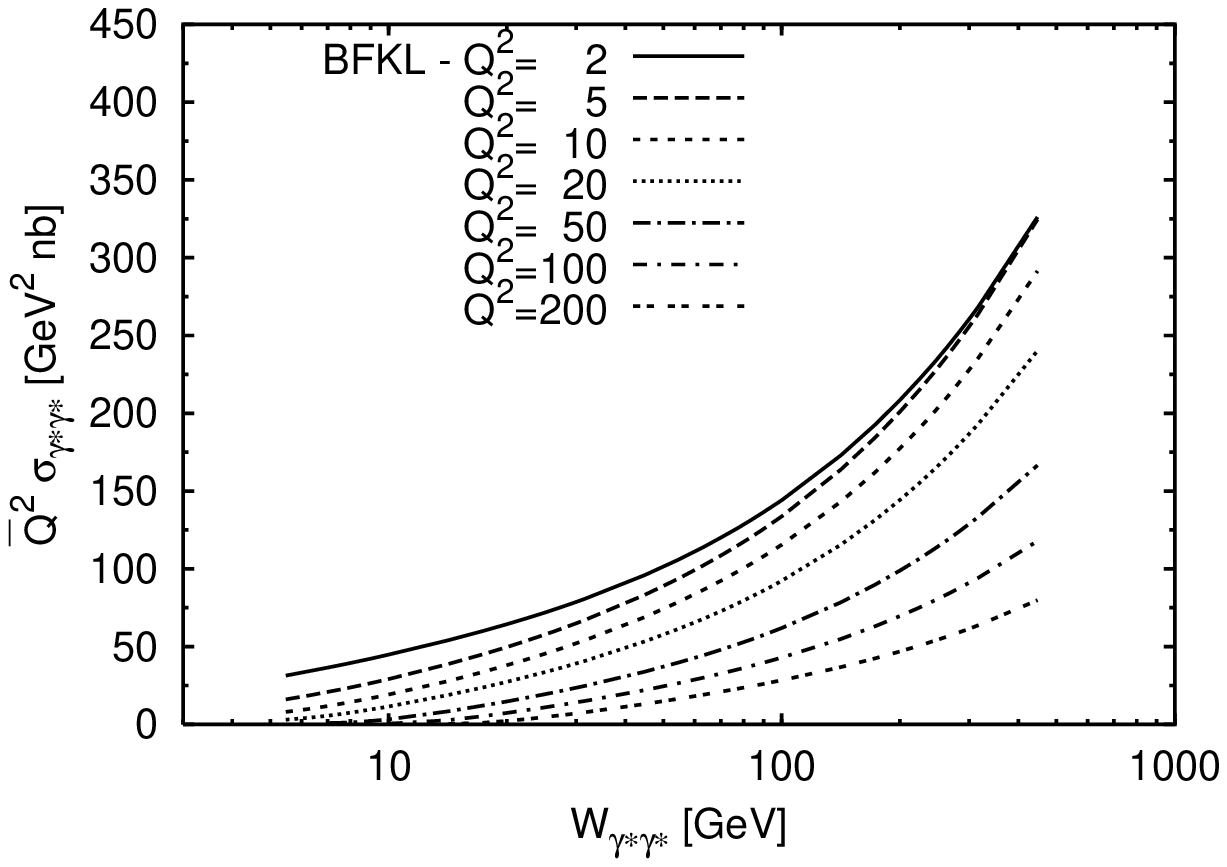} \\
c) \\
\caption{\small\it Predictions for the re-scaled gluonic component of the two virtual photon
cross-section $\bar{Q} ^2 \tilde\sigma^G _{\gamma^*\gamma^*}(W^2,Q^2,Q^2)$ obtained
using the parameters of Model~1 for (a)  $Q^2$ between 0~and 10~GeV$^2$
and (b) between 10 and 200~GeV$^2$, and (c) the same quantity from
the BFKL solution combined with the soft pomeron contribution.}
\label{future-gg}
\end{center}
\end{figure}

For comparison the same quantity, but obtained from the solution of the BFKL equation
with subleading corrections in the perturbative domain \cite{JKLM},
is presented in Fig.~\ref{future-gg}c. Note, that the latter
result contains not only the perturbative BFKL part but also
the soft pomeron contribution, obtained using the Regge factorisation.
Recall that the relative importance of the soft pomeron term quickly decreases
with the increasing photon virtuality $Q^2$.

It is interesting to observe, that in the saturation model, at higher
$W$ where the threshold corrections are negligible, the quantity
$\bar{Q} ^2 \tilde\sigma^G_{\gamma^*\gamma^*}(W^2,Q^2,Q^2)$ increases with $Q^2$
up to about 10~GeV$^2$, see  Fig.~\ref{future-gg}a and for
$Q^2>10$~GeV$^2$ it decreases with $Q^2$. The reason for the
increase for smaller $Q^2$ is the rising contribution from
the charmed quark dipoles. The relative suppression of the charmed
quark contribution at $Q^2 = 0$, in comparison to the light quarks
due to the higher charm mass, is becoming less important towards higher
$Q^2 > 4m_c ^2$, when the typical scales in those two cases
become similar.

On the other hand, the bottom quark has a relatively small charge
of $1/3\, e$ and such threshold effects are much less pronounced.
Thus, for $Q^2 > 4m_c ^2$ the cross-section should enter the
geometric scaling regime. The unitarity corrections may be neglected
at higher $Q^2$ and it follows that
$\bar{Q} ^2 \tilde\sigma^G_{\gamma^*\gamma^*}(W^2,Q^2,Q^2) \sim (W^2/Q^2)^{\lambda}$,
modulo threshold corrections. This is the reason  why one sees a monotonical decrease of
$\bar{Q} ^2 \tilde\sigma^G_{\gamma^*\gamma^*}(W^2,Q^2,Q^2)$ with $Q^2$ in
Fig.~\ref{future-gg}b.

For $Q^2$ in the perturbative domain, the difference between the results
from saturation model and the BFKL predictions is not large but grows with $Q^2$.
The tendency of $\bar{Q} ^2 \tilde\sigma^G _{\gamma^*\gamma^*}(W^2,Q^2,Q^2)$
to decrease with increasing $Q^2$ in the BFKL approach may be
traced back to important threshold effects and the running of the QCD coupling,
which were taken into account in \cite{JKLM}.

\begin{figure}[t]
\begin{center}
\epsfig{width= 0.6\columnwidth,file=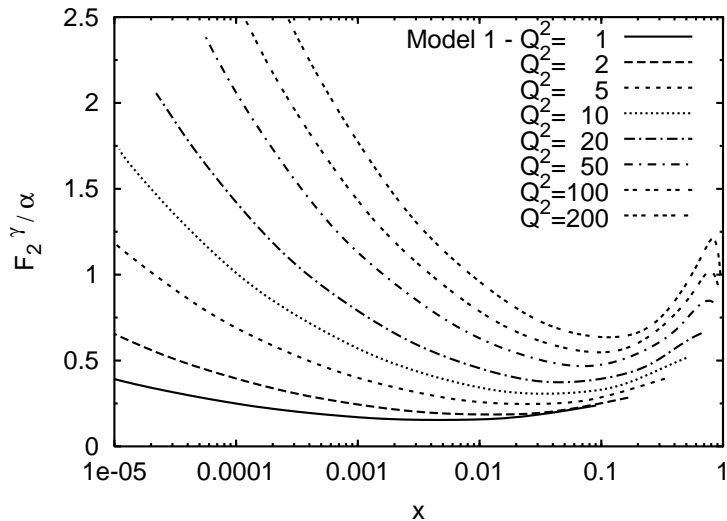}
\caption{\small\it The real photon structure function $F_2 ^{\gamma}(x,Q^2)$
for various $Q^2$ from Model 1.}
\label{future-f2}
\end{center}
\end{figure}

Let us also recall that the prediction for the $\gamma\gamma$ total cross-section
for $W$ in the TeV range is not sensitive to the choice of the form of the
dipole-dipole cross-section, see Fig.~\ref{real-all}. However, it does rely on the
accuracy of the data unfolding with \textsc{Phojet}. Thus, the systematic uncertainty
of the data unfolding at LEP propagates into the model predictions.

In Fig.~\ref{future-f2} the real photon structure $F_2^{\gamma} (x,Q^2)$ is plotted
for various $Q^2$ between 1~GeV$^2$ and 200~GeV$^2$ for $x$ down to $10^{-5}$.
The kinematical range was chosen to be relevant for the future linear collider
measurements.
For completeness, in Fig.~\ref{future-charm} we also give the dependence of the
cross-section for heavy quarks production in $\gamma\gamma$ collisions
in a wide range of two-photon collision energy~$W$.
We indicate the effect of the quark mass variation both for
charm and for bottom.  One should, however, keep in mind, that the saturation model
in the present form gives slightly too low cross-sections for $c$~quarks production and
significantly too low (by factor of about 2 -- 2.5) for $b$~quarks.

\begin{figure}[t]
\begin{center}
\epsfig{width= 0.6\columnwidth,file=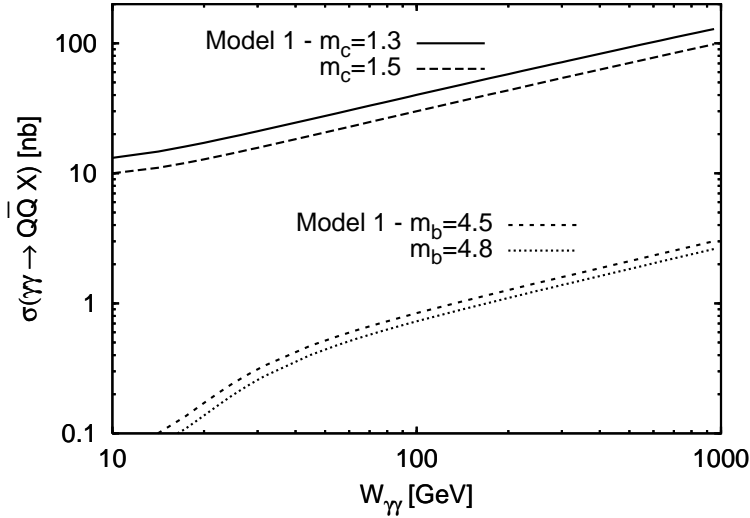}
\caption{\small\it Cross-sections for heavy quark production in two real photon collisions
from Model~1.}
\label{future-charm}
\end{center}
\end{figure}

\section{Conclusions}

In this paper we have extended the saturation model proposed by
Golec-Biernat and W\"{u}sthoff for $\gamma^* p$ cross-sections,
to describe two-photon processes at high energies.
This extension required an explicit model for the scattering of
two colour dipoles. We considered three models of this cross-section,
all of them exhibiting the essential feature of colour transparency
for small dipoles, and the saturation property for large ones.
We kept the GBW form of the unitarising function and the original
parameters, except for changing the values of quark masses, which 
was necessary to describe the data on the 
total two real photon cross-section. 
We have explicitly checked that these modifications
had not spoiled the fit to photoproduction data.
In order to obtain a more complete description applicable at lower energies
the saturation model has been combined with other, well known contributions  
related to the quark box diagram and non-pomeron reggeon exchange. 
Those mechanisms become dominant when the collision energy is 
comparable to photon virtualities and quark masses. 
Standard multiplicative threshold correction factors,  
relevant for lower energies, 
have been included into the saturation model and the sub-leading reggeon 
contribution.

We have analysed  general features of the saturation model for the $\gamma\gamma$
scattering. We observed, that such models gave the energy ($W$) dependence
for two-photon total cross-section in the saturation regime steeper
than the photon-hadron and hadron-hadron total cross-sections, by additional
factors of $\ln W$ and $\ln^2 W$ respectively. In the non-saturated regime,
a typical power-like growth $W^{2\lambda}$, $\lambda \simeq 0.3$, with energy was obtained, 
characteristic for the hard pomeron exchange.

The results from the studied models were compared with the data for
different two-photon processes at high rapidity values: the total
$\gamma\gamma$ cross-section, the total $\gamma^*\gamma^*$ cross-section
for similar virtualities of the photons, the real photon structure
function $F_2 ^{\gamma}$ and heavy flavour production. Free parameters
of the models were fitted to the data. It was found, that the data favour
one of the models for dipole-dipole cross-section, namely Model~1 presented in 
Section 2.2 and Section 3.1.
With this model a reasonable global description of the available two-photon
data was obtained, except for the $b$-quark production.
Predictions for energies accessible at future linear colliders
were formulated.
It is however encouraging, that Model~2 (see Sec.\ 2.2 and Sec.\ 3.1),
being a significantly different generalisation of the original GBW model, 
gives the results close to Model~1, with a relative difference of
less than 15\%. This means that the sensitivity of the predictions from the 
saturation model to the details is not very significant.

In summary, the saturation model was found to provide a simple and efficient
framework to calculate observables in two-photon processes. This success
supports strongly the idea of rapidity dependent saturation scale and
improves the understanding of two-photon physics.

\section*{Acknowledgments}
We are indebted to Rikard Enberg for his contribution at the early stage
of this work and his useful remarks.
We thank
Barbara Bade\l{}ek and
Gunnar Ingelman for reading the manuscript and
their comments.
LM is grateful to the Swedish Natural Science Research Council
for the fellowship.
This research was partially supported
by the EU Fourth Framework Programme `Training and Mobility of Researchers',
Network `Quantum Chromodynamics and the Deep Structure of Elementary
Particles', contract FMRX--CT98--0194 and  by the Polish
Committee for Scientific Research (KBN) grants no. 2P03B 05119 and 5P03B 14420.


\newpage

\begin{thebibliography}{9999}
\bibitem{GLR} L.~V.~Gribov, E.~M.~Levin and M.~G.~Ryskin, Phys.\ Rep.\ {\bf 100} (1983) 1.

\bibitem{MQ} A.~H.~Mueller and J.~Qiu, Nucl.\ Phys.\ {\bf B268} (1986) 427.

\bibitem{NIKO} N.~N.~Nikolaev and B.~G.~Zakharov,  Z.\ Phys.\ {\bf C49} (1991) 607;
Z.\ Phys.\ {\bf C53} (1992) 331;  Z.\ Phys.\ {\bf C64} (1994) 651;
JETP {\bf 78} (1994) 598.

%
\bibitem{MUELLER} A.H.~Mueller, Nucl. Phys.  {\bf B415} (1994) 373;
A.~H.~Mueller and B.~Patel, Nucl.\ Phys.\ {\bf B425} (1994) 471;
A.~H.~Mueller, Nucl.\ Phys.\ {\bf B437} (1995) 107.


%
\bibitem{MUELLERS}
A.~H.~Mueller,  Nucl.\ Phys.\  {\bf B335} (1990) 115;
Yu.~A.~Kovchegov, A.~H.~Mueller and S.~Wallon,  Nucl.\ Phys.\ {\bf B507} (1997) 367.
A.~H.~Mueller,  Eur.\ Phys.\ J.\ {\bf A1} (1998) 19;
 Nucl.\ Phys.\ {\bf  A654} (1999) 370;
 Nucl.\ Phys.\ {\bf B558}  (1999) 285.

\bibitem{COLLINS} J.~C.~Collins and J.~Kwieci\'nski,  Nucl.\ Phys.\  {\bf B335} (1990) 89;
J.~Bartels, G.~A.~Schuler and  J.~Bl\"umlein,  Z.\ Phys.\ {\bf C50} (1991) 91;
Nucl.\ Phys.\ Proc.\ Suppl.\  {\bf  18 C} (1991) 147.
\bibitem{BARTELS} J.~Bartels and E.~M.~Levin,  Nucl.\ Phys.\ {\bf B387} (1992) 617;
 J.~Bartels,  Phys.\ Lett.\ {\bf B298} (1993) 204;
      Z.\ Phys.\ {\bf C60} (1993) 471; Z.\ Phys.\  {\bf C62} (1994) 425;
 J.~Bartels and M.~W\"usthoff, Z.\ Phys.\ {\bf C66} (1995) 157;
 J.~Bartels and C.~Ewerz, JHEP {\bf 9909} (1999) 026.

%
\bibitem{VENUGOPALAN} L.~McLerran and  R.~Venugopalan,
Phys.\ Rev.\ {\bf D49} (1994) 2233;
Phys.\ Rev.\ {\bf D49} (1994) 3352;
Phys.\ Rev.\ {\bf D50} (1994) 2225;
A.~Kovner, L.~McLerran and H.~Weigert, Phys.\ Rev.\ {\bf D52} (1995) 6231,
Phys.\ Rev.\ {\bf D52} (1995) 3809;
R.~Venugopalan,  Acta Phys.\ Polon.\ {\bf B30} (1999) 3731;
E.~Iancu and L.~McLerran, Phys.\ Lett.\ {\bf B510} (2001) 145;
L.~McLerran, {hep-ph/0104285};
E.~Iancu, A.~Leonidov and L.~McLerran, Nucl.\ Phys.\ {\bf A692} (2001) 583;
E.~Ferreiro, E.~Iancu, A.~Leonidov and L. McLerran, {hep-ph/0109115};
A.~Capella {\it et al.}, Phys.\ Rev.\ {\bf D63} (2001) 054010.



\bibitem{SALAM} G.~P.~Salam,  Nucl.\ Phys.\ {\bf B449} (1995) 589;
Nucl.\ Phys.\  {\bf B461} (1996) 512;
Comput.\ Phys.\ Commun.\ {\bf 105} (1997) 62;
A.~H.~Mueller and  G.~P.~Salam, Nucl.\ Phys.\  {\bf B475} (1996) 293.


\bibitem{LEVIN} E.~Gotsman, E.~M.~Levin and  U.~Maor,
    Nucl.\ Phys.\ {\bf B464} (1996) 251;
    Nucl.\ Phys.\ {\bf B493} (1997) 354;
    Phys.\ Lett.\ {\bf B245} (1998) 369;
    Eur.\ Phys.\ J.\ {\bf C5} (1998) 303;
    E.~Gotsman, E.~M.~Levin,  U.~Maor and E.~Naftali,
    Nucl.\ Phys.\ {\bf B539} (1999) 535;
    A.~L.~Ayala~Filho, M.~B.~Gay~Ducati and  E.~M.~Levin, Nucl.\ Phys.\ {\bf B493}
    (1997) 305; Nucl.\ Phys.\ {\bf B551} (1998) 355;
    Eur.\ Phys.\ J.\ {\bf C8} (1999) 115.


\bibitem{BAL} Ia.~Balitsky, Nucl.\ Phys.\ {\bf B463}  (1996) 99.

\bibitem{WEIGERT}  J.~Jalilian-Marian, A.~Kovner, L.~McLerran  and  H.~Weigert,
Phys.\ Rev.\ {\bf D55} (1997) {5414};
J.~Jalilian-Marian, A.~Kovner and  H.~Weigert, Phys.\ Rev.\ {\bf D59} (1999) {014014};
Phys.\ Rev.\ {\bf D59} (1999) 014015;
Phys.\ Rev.\ {\bf D59} (1999) 034007; Erratum -- ibid.\ {\bf D59} (1999) 099903;
A.~Kovner, J.~Guilherme~Milhano and  H.~Weigert,
Phys.\ Rev.\ {\bf D62} (2000) 114005;
H.~Weigert, NORDITA-2000-34-HE, {hep-ph/0004044}.


\bibitem{BRAUN}
M.~A.~Braun, Eur.\ Phys.\ J.\ {\bf C16} (2000) 337; {hep-ph/0101070}.

\bibitem{K} Yu.~V.~Kovchegov and L.~McLerran,
Phys.\ Rev.\ {\bf D60} (1999) 054025; Erratum -- ibid.\ {\bf D62} (2000) 019901;
Yu.~V.~Kovchegov and E.~M.~Levin, Nucl.\ Phys.\ {\bf B577} (2000) 221;
Yu.~V.~Kovchegov, Phys.\ Rev.\ {\bf D60} (1999) 034008;
Phys.\ Rev.\ {\bf D61} (2000) 074018.

\bibitem{KSOL}
E.~M.~Levin and K.~Tuchin, Nucl.\ Phys.\ {\bf B537} (2000) 833;
Nucl.\ Phys.\ {\bf A691} (2001) 779; ibid.\ {\bf A693} (2001) 787.

%
\bibitem{KGBMW} K.~Golec-Biernat and M.~W\"usthoff, Phys.\ Rev.\ {\bf D59} (1998) 014017.

\bibitem{BFKL} E.~A.~Kuraev, L.~N.~Lipatov and V.~S.~Fadin, Sov.\ Phys.\ JETP
{\bf 44} (1976) 443; ibid.\ {\bf 45} (1977) 199;
I.~I.~Balitsky and L.~N.~Lipatov, Sov.\ J.\ Nucl.\ Phys.\ {\bf 28} (1978) 822.

\bibitem{MANDY} A.~M.~Cooper-Sarkar, R.C.E. Devenish, A. De Roeck,  
Int.\ J.\ Mod.\ Phys.\ {\bf C7} (1999) 609.

\bibitem{KGBMWD} K.~Golec-Biernat and M.~W\"usthoff, Phys.\ Rev.\ {\bf D60} (1999) 114023.
\bibitem{GLM} E.~Gotsman, A.~Levy and U.~Maor, Z.\ Phys.\ {\bf C40} (1988) 117.
\bibitem{GALUGA} G.~A.~Schuler, Comput.\ Phys.\ Commun.\ {\bf 108} (1998) 279.
%
%
\bibitem{SSGG} G.~A.~Schuler and T.~Sj\"ostrand, Z.\ Phys.\ {\bf C73} (1997) 677.
%
\bibitem{DOSCH} A.~Donnachie, H.~G.~Dosch and M.~Rueter, Phys.\ Rev.\ {\bf D59} (1999) 074011;
  H.~G.~Dosch, Nucl.\ Phys.\ Proc.\ Suppl.\ {\bf 96} (2001) 118.
\bibitem{DDR} A.~Donnachie, H.~G.~Dosch and M.~Rueter, Eur.\ Phys.\ J.\ {\bf C13} (2000) 141.
\bibitem{KOLYA}N.~N.~Nikolaev, J.~Speth and  V.~R.~Zoller,  hep-ph/0001120.
%
\bibitem{GOTSMAN} E.~Gotsman {\it et al.\ },
Eur.\ Phys.\ J.\ {\bf C14} (2000) 511.
%
\bibitem{BKS}
B.~Bade\l{}ek, J.~Kwieci\'{n}ski and A.~M.~Sta\'{s}to,
Acta Phys.\ Polon.\ {\bf B30} (1999) 1807.
\bibitem{BKKS}
B.~Bade\l{}ek, M.~Krawczyk, J.~Kwieci\'{n}ski and A.~M.~Sta\'{s}to,
Phys.\ Rev.\ {\bf D62} (2000) 074021.
\bibitem{FS}  J.~R.~Forshaw and J.~K.~Storrow, Phys.\ Rev.\ {\bf D46} (1992) 4955;
Phys.\ Lett.\ {\bf B278} (1992) 193.
\bibitem{CGP} A.~Corsetti, R.~M.~Godbole and G.~Pancheri, Phys.\ Lett.\ {\bf B435}
(1998) 441 and references therein;
R.~M.~Godbole, A.~Grau and G.~Pancheri, Nucl.\ Phys.\ Proc.\ Suppl.\
{\bf 82} (2000) 246; R.~M.~ Goodbole and  G.~Pancheri,
Nucl.\ Instrum.\ and Meth.\ {\bf A472} (2001) 205.
%
%
\bibitem{BBJKRMP} B. Bade\l{}ek, J. Kwieci\'nski, Rev.\ Mod.\ Phys.\ {\bf 68} (1996) 445.
\bibitem{KTFAC}S.~Catani, M.~Ciafaloni and F.~Hautmann, Phys.\ Lett.\ {\bf B242} (1990) 97;
Nucl.\ Phys.\ {\bf B366} (1991) 135;
J.~C.~Collins and R.~K.~Ellis, Nucl.\ Phys.\ {\bf B360} (1991) 3;
S.~Catani and F.~Hautmann, Nucl.\ Phys.\ {\bf B427} (1991) 475.
%
\bibitem{BIALAS} A.~Bia\l{}as, H.~Navelet and R.~Peschanski, Nucl.\ Phys.\ {\bf B593} (2001) 438.
%
\bibitem{TWOP} A.~Donnachie and P.~V.~Landshoff, Phys.\ Lett.\ {\bf B437} (1998) 408.
%
\bibitem{GSGSBFKL}
S.~J.~Brodsky, F.~Hautmann and D.~A.~Soper,  Phys.\ Rev.\ {\bf D56} (1997) 6957;
Phys.\ Rev.\ Lett.\ {\bf 78} (1997) 803; Erratum -- ibid.\ {\bf 79} (1997) 3544;
J.~Bartels, A.~De~Roeck and H.~Lotter, Phys.\ Lett.\  {\bf B389} (1996) 742;
J.~Bartels, A.~De Roeck, C.~Ewerz and H.~Lotter, hep-ph/9710500;
W.~Florkowski, Acta Phys.\ Polon.\ {\bf B28} (1997) 2673;
A.~Bia\l{}as, W. Czy\.z and W. Florkowski, Eur.\ Phys.\ J.\ {\bf C2}  (1998) 683;
M.~Boonekamp {\it et al.}, Nucl.\ Phys.\ {\bf B555} (1999) 540;
V.~T.~Kim, L.~Lipatov and G.~B.~Pivovarov, hep-ph/9911228;
J.~Bartels, C.~Ewerz and R.~Staritzbichler, Phys.\ Lett.\ {\bf B492} (2000) 56.
%
\bibitem{JKLM}
J.~Kwieci\'{n}ski and L.~Motyka,
Acta Phys.\ Polon.\ {\bf B30} (1999) 1817;
Phys.\ Lett.\ B {\bf 462} (1999) 203;
Eur.\ Phys.\ J.\ {\bf C18} (2000) 343.
%
\bibitem{GEOMETRIC} A.~Sta\'sto, K.~Golec-Biernat and J.~Kwieci\'nski,
Phys.\ Rev.\ Lett.\ {\bf 86} (2001) 56.
%
\bibitem{BUD} V.~M.~Budnev, I.~F.~Ginzburg, C.~V.~Meledin and V.~G.~Serbo,
Phys.\ Rep.\ {\bf 15} (1974) 181.
%
\bibitem{DLspec}
A.~Donnachie and P.~V.~Landshoff,
Phys.\ Lett.\ {\bf B518} (2001) 63.
%
\bibitem{PVLF0}P.~V.~Landshoff, hep-ph/0010315.
%
\bibitem{KZS}
M.~Krawczyk, A.~Zembrzuski and M.~Staszel,
Phys.\ Rep.\  {\bf 345} (2001) 265.
%
\bibitem{DLgp}
A.~Donnachie and P.~V.~Landshoff,
Phys.\ Lett.\ {\bf B296} (1992) 227;

Z.\ Phys.\ {\bf C61} (1994) 139.

\bibitem{PDG}
D.~E.~Groom {\it et al.}  [Particle Data Group Collaboration],
Eur.\ Phys.\ J.\ {\bf C15} (2000) 1.

\bibitem{GGTOTEXP1}
C.~Berger {\it et al.}  [PLUTO Collaboration],
Z.\ Phys.\ {\bf C26} (1984) 353;
C.~Berger {\it et al.}  [PLUTO Collaboration],
Phys.\ Lett.\ {\bf B149} (1984) 421;
%
D.~Bintinger {\it et al.}  [TPC/Two Gamma Collaboration],
Phys.\ Rev.\ Lett.\  {\bf 54} (1985) 763.
H.~Aihara {\it et al.}  [TPC/Two Gamma Collaboration],
Phys.\ Rev.\ {\bf D41} (1990) 2667;
%
%
S.~E.~Baru {\it et al.},
Z.\ Phys.\ {\bf C53} (1992) 219;


\bibitem{GGTOTEXP2}
%
%
G.~Abbiendi {\it et al.}  [OPAL Collaboration],
Eur.\ Phys.\ J.\ {\bf C14} (2000) 199.
%
%
M.~Acciarri {\it et al.}  [L3 Collaboration],
Phys.\ Lett.\ {\bf B519} (2001) 33;
%


\bibitem{pythia}
T.~Sj\"{o}strand, P.~Eden, C.~Friberg, L.~L\"{o}nnblad, G.~Miu, S.~Mrenna and E.~Norrbin,
Comput.\ Phys.\ Commun.\  {\bf 135} (2001) 238.

\bibitem{phojet}
R.~Engel and J.~Ranft,
Phys.\ Rev.\ {\bf D54} (1996) 4244.




\bibitem{DTL3}
M.~Acciarri {\it et al.}  [L3 Collaboration],
Phys.\ Lett.\ {\bf B453} (1999) 333;
L3 Note 2680 (contribution to EPS-HEP conference in Budapest, July 2001).


\bibitem{DTOPAL}
G.~Abbiendi {\it et al.}  [OPAL Collaboration],
hep-ex/0110006.




\bibitem{F2EXP1}
%
%
C.~Berger {\it et al.}  [PLUTO Collaboration],
Phys.\ Lett.\ {\bf B142} (1984) 111;
%
%
%
%
H.~Aihara {\it et al.}  [TPC/Two Gamma Collaboration],
Z.\ Phys.\ {\bf C34} (1987) 1;
%
%
%
C.~Berger {\it et al.}  [PLUTO Collaboration],
Nucl.\ Phys.\ {\bf B281} (1987) 365;
%
%
%
K.~Muramatsu {\it et al.}  [TOPAZ Collaboration],
Phys.\ Lett.\ {\bf B332} (1994) 477.

\bibitem{F2EXP2}
M.~Acciarri {\it et al.}  [L3 Collaboration],
Phys.\ Lett.\ {\bf B436} (1998) 403;
%
%
M.~Acciarri {\it et al.}  [L3 Collaboration],
Phys.\ Lett.\ {\bf B447} (1999) 147;
%
%
G.~Abbiendi {\it et al.}  [OPAL Collaboration],
Eur.\ Phys.\ J.\ {\bf C18} (2000) 15;
%
%
R.~Barate {\it et al.}  [ALEPH Collaboration],
Phys.\ Lett.\ {\bf B458} (1999) 152;
%
%
P.~Abreu {\it et al.}  [DELPHI Collaboration],
Z.\ Phys.\ {\bf C69} (1996) 223.



\bibitem{L3cc}
M.~Acciarri {\it et al.}  [L3 Collaboration],
Phys.\ Lett.\ {\bf B514} (2001) 19.


\bibitem{L3bb}
M.~Acciarri {\it et al.}  [L3 Collaboration],
Phys.\ Lett.\ {\bf B503} (2001) 10.
%
\bibitem{OPbb}
The OPAL Collaboration, OPAL Physics Note PN455.
%

\bibitem{TESLA}
J.~A.~Aguilar-Saavedra {\it et al.}  [ECFA/DESY LC Physics Working Group
                  Collaboration],
hep-ph/0106315; 
B.~Bade\l{}ek {\it et al.}  [TESLA-N Study Group Collaboration],
Nucl.\ Instr. Meth.\ {\bf A472} (2001) 1 [hep-ex/0108012].



\end{thebibliography}
\end{document}